\documentclass[11pt]{article}

\usepackage[utf8]{inputenc}
\usepackage[T1]{fontenc}
\usepackage{lmodern}
\usepackage[margin=1in]{geometry}
\usepackage{amsmath,amssymb}
\usepackage{bm}
\usepackage{graphicx}
\usepackage[labelfont=bf]{caption}
\usepackage[numbers,sort&compress]{natbib}
\usepackage{authblk}
\usepackage{microtype}
\usepackage[colorlinks=true,allcolors=blue]{hyperref}
\renewcommand{\eqref}[1]{\textup{Eq.~(\ref{#1})}}

\begin{document}

\title{Polarization of the \textit{Drosophila} oocyte by active cortical exchange}

\author[a,*]{Mohammadhossein Firouznia}
\author[b]{Wen Lu}
\author[b]{Margot Lakonishok}
\author[a]{Reza Farhadifar}
\author[b]{Vladimir~I.~Gelfand}
\author[a,c,*]{Michael~J.~Shelley}

\affil[a]{Center for Computational Biology, Flatiron Institute, New York, NY 10010}
\affil[b]{Department of Cell and Developmental Biology, Feinberg School of Medicine, Northwestern University, Chicago, IL 60611}
\affil[c]{Courant Institute, New York University, New York, NY 10012}
\affil[*]{To whom correspondence should be addressed. E-mail:\,hfirouznia@flatironinstitute.org, mshelley@flatironinstitute.org}

\date{}

\maketitle

\begin{abstract}

Anterior--posterior polarization in the \textit{Drosophila} oocyte requires posterior localization of \textit{oskar} mRNA. \textit{oskar} mRNA is produced in the anteriorly situated nurse cells, moved into the oocyte, and assembled into \textit{oskar}--Staufen ribonucleoprotein particles (RNPs). These RNPs then move through the oocyte and interact with the cortex, where myosin-V binds them to the actin network and kinesin-1 removes them along cortically anchored microtubules. This cortical tug-of-war takes place against an anterior-to-posterior decrease in microtubule density. How the interplay between cytoplasmic transport and cortical binding and unbinding produces a stable posterior cap is unclear. Here, we combine quantitative microscopy, photoconversion experiments, experimental perturbations, and biophysical modeling to show that posterior localization emerges from position-dependent cortical exchange coupled to cytoplasmic diffusion. Our model, which accounts for the tendency of RNPs to demix, suggests that two key parameters that determine the observed localization pattern are the total amount of \textit{oskar} and the ratio of cortical unbinding to binding rates (the exchange equilibrium constant). Given \textit{oskar} volume fractions in wild-type, the measured exchange constant is such that \textit{oskar} localizes to the posterior cortex without demixing. If the balance is broken, say by increasing unbinding rates, \textit{oskar} irreversibly demixes into cytoplasmic droplets. This is consistent with observations from our experimental perturbations. Our results show how bulk diffusion and cortical exchange, in the face of \textit{oskar} RNPs' self-attraction, convert a cytoskeletal asymmetry into robust developmental polarity.

\end{abstract} %


\section*{Introduction}

Cells organize their contents in space to carry out essential biological functions. This organization is achieved through a variety of physical mechanisms operating across different structures. For example, the mitotic spindle assembles into a bipolar array through the nucleation of microtubules from centrosomes and chromosomes, which then positions the spindle within the cell and segregates chromosomes \cite{heald1996self, grill2003distribution, 10.7554/eLife.55877}. Within the nucleus, the genome is partitioned into compact domains, where transcription is suppressed, and loosely packed domains, where gene transcription is more active. This compaction is facilitated by DNA methylation, which promotes the binding of crosslinking proteins to chromatin \cite{cheutin2003maintenance,mahajan2022euchromatin,lamson2025condensation,rautu2026active}. Membraneless organelles such as P-granules form when their protein constituents phase separate from the surrounding cytoplasm, concentrating germline determinants that specify the germ cell lineage \cite{brangwynne2009germline, hymanLiquidLiquidPhaseSeparation2014, Weber_2019}. Biophysical modeling has been essential for understanding the principles that govern this spatial organization.

The \textit{Drosophila} oocyte is a well-established model organism for studying how such spatial patterns are established during development. Its anterior-posterior (AP) axis is set by a group of maternal mRNAs, including \textit{bicoid}, \textit{nanos}, and \textit{oskar}, that localize to distinct regions of the oocyte and break its initial uniformity \cite{chekulaeva2024mechanistic}. The \textit{oskar} mRNA provides one of the best-characterized examples. It is transcribed in the nurse cells and actively transported into the oocyte, by dynein motors through interconnected cytoplasmic bridges called ring canals \cite{RN9,RN10,RN11,RN12}. \textit{Oskar} then associates with the double-stranded RNA-binding protein Staufen to form ribonucleoprotein particles (RNPs) that serve as the functional units of transport and localization \cite{RN4,RN5,RN6}. At stage 7, early in oogenesis, \textit{oskar} is distributed throughout the cytoplasm, whereas by the end of stage 9 it is concentrated in the posterior cap. This \textit{oskar}--Staufen cap then specifies the site of pole plasm assembly. It is known that \textit{oskar} localization requires microtubules and the plus-end directed kinesin-1 motor \cite{RN21}. %

A previous study by Khuc Trong \textit{et al.} proposed that the transport of \textit{oskar} RNPs along cortical microtubules, without global order, together with an anterior-posterior gradient in microtubule density, can explain their localization at the posterior of the oocyte \cite{RN22}. However, new experiments have given new information about how suspended RNPs interact with the cortex, and how they interact amongst themselves. Building on previous measurements, we show that there is an active exchange of RNPs between the bulk cytoplasm and the cortex that is sculpted by the gradient in cortical microtubule density \cite{RN24}. This exchange is driven by the competition between kinesin-1 motor removing \textit{oskar} from the cortex, and the myosin-V motor binding it to the cortex. We hypothesize that it is this sculpted exchange that drives RNP localization to the posterior. We further show, as suggested by previous observations \cite{bose_Cell_2022_liquid-to-solid_phase_transition}, that there is a tendency towards aggregation of suspended RNPs, suggesting that the wild-type cortical exchange mechanism may be tuned to keep the bulk RNP concentration below
a phase segregation threshold.

Here, we present new experiments and a model that accounts for active exchange of \textit{oskar} particles between the bulk cytoplasm and the cortex, driven by kinesin-1 and myosin-V motors, and the self-interaction of \textit{oskar}--Staufen RNPs in the bulk. We combine theory and experiments to quantitatively measure all the biophysical parameters of the model, such as the profile of binding and unbinding rates of RNPs along the cortex, the cytoplasmic volume fraction of RNPs, and the parameters characterizing \textit{oskar}--Staufen RNPs' self-interaction. We show that this model explains the patterning dynamics of \textit{oskar} in wild-type and mutant stage 7--9 oocytes as an interplay between spatially regulated binding and unbinding at the oocyte cortex and phase separation in the bulk.   

\begin{figure}%
\centering
\includegraphics[width=0.55\linewidth]{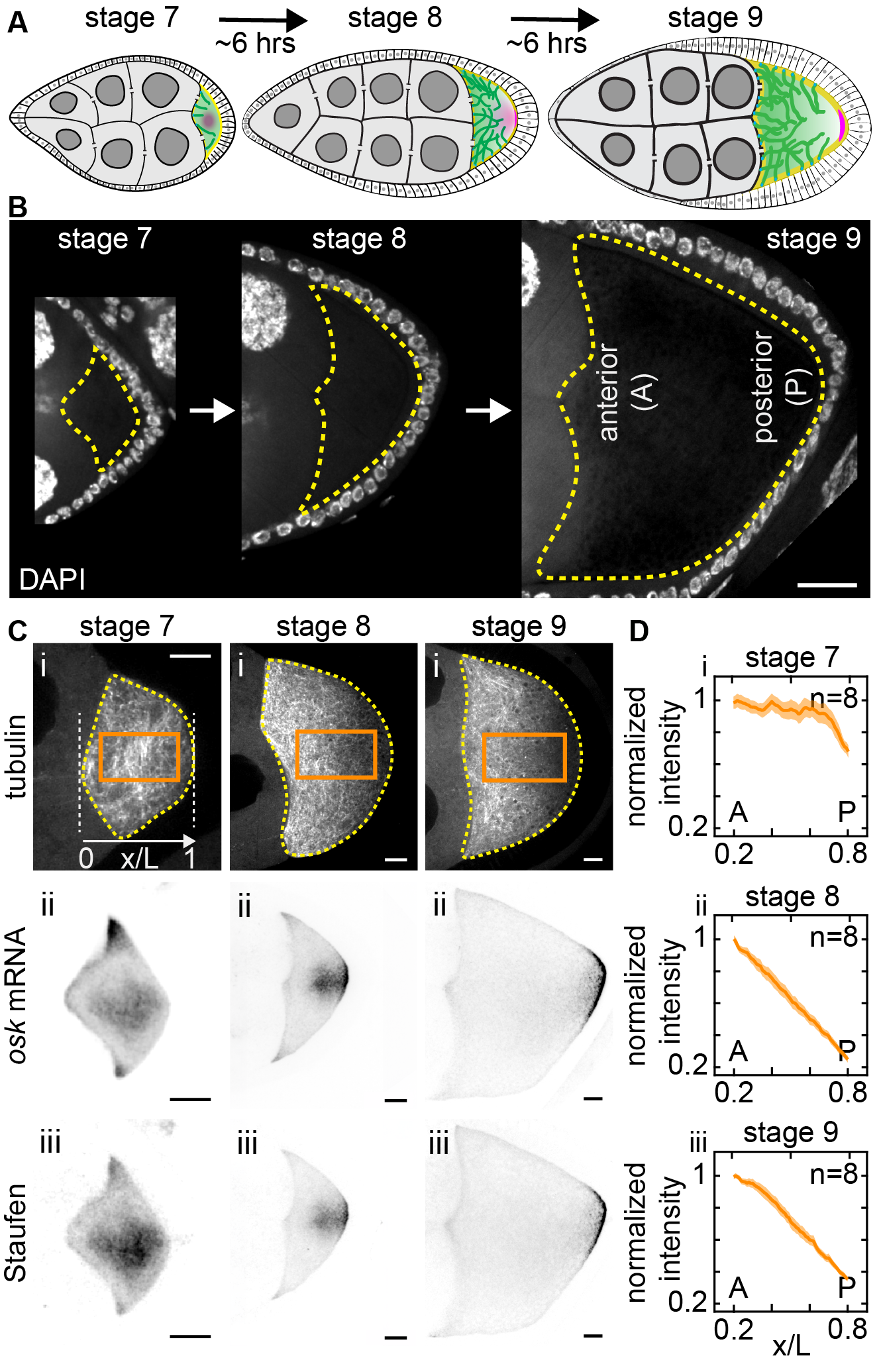}
\caption{Dynamic reorganization of microtubules and \textit{oskar}--Staufen RNPs in growing \textit{Drosophila} oocytes. (A) Schematic of  \textit{Drosophila} egg chambers at stages 7, 8, and 9. The nurse cells (gray), oocyte  (yellow), microtubules (green), and \textit{oskar} particles (magenta) are shown. The times between developmental stages are indicated. (B) Mid-plane sections of DAPI-stained oocytes at stages 7, 8, and 9. The yellow dashed line shows the outline of the oocyte. The anterior (A) and posterior (P) sides of the oocytes are indicated. Scale bar, $20$ $\mu\mathrm{m}$. (C) Microtubule, \textit{oskar}, and Staufen distributions during oocyte development: Snapshots of far-red tubulin imaged at the cortex in stages 7, 8, and 9 oocytes indicating microtubule organization, is shown in (i). Normalized distance from the anterior side is defined as $x/L$ ($=0$ at the anterior side). The yellow dashed line shows the outline of the oocyte, and the white dashed lines correspond to $x/L=0,1$. The orange box shows the region for measurements of the tubulin intensity. \textit{oskar} mRNA distribution in the mid-plane of stage 7, 8, and 9 oocytes are shown in (ii). Staufen distribution in the mid-plane of stage 7, 8, and 9 oocytes are shown in (iii). Scale bars in panels (i) to (iii) is $10$ $\mu\mathrm{m}$ . (D) Normalized tubulin intensity measured in the orange box shown in (C-i) is plotted as a function of $x/L$, by averaging the intensity perpendicular to the AP axis (n=8). }
\label{fig:osk_staufen_stage789}
\end{figure}

\section*{Results}
\subsection*{Formation of microtubule gradient and \textit{oskar} localization in \textit{Drosophila} oocyte}

During \textit{Drosophila} oogenesis, from stage 7 to 9, the oocyte undergoes substantial changes: its volume increases, microtubule organization changes, and different molecules localize to different parts of the oocyte (Fig.~\ref{fig:osk_staufen_stage789}A). We quantified the changes in oocyte size by fixing n=18 egg chambers at stages 7--9 and staining the cell periphery with phalloidin (that labels F-actin) and the nuclei with DAPI (that labels double-stranded DNA) to distinguish the oocyte from the follicle cells forming the outer layer of the egg chamber (see Fig.~S1 in SI). We measured the oocyte length as the distance from the anterior side (shared with the nurse cells) to the posterior pole in the mid-plane of the egg chamber and found that the length increases from $27\pm 5~\mathrm{\mu m}$ to $77\pm 21~\mathrm{\mu m}$ during $\sim12$ hours of oogenesis (Fig.~\ref{fig:osk_staufen_stage789}B). The growth is predominantly from the anterior side \cite{RN26}. 

We visualized the microtubule organization in stage 7 to 9 oocytes using InfraRed-$\mathrm{\alpha}$tub in live cells (Fig.~\ref{fig:osk_staufen_stage789}\mbox{C-i}) and measured the average tubulin intensity along the AP axis. During stage 7, tubulin is uniformly distributed with little apparent AP bias (Fig.~\ref{fig:osk_staufen_stage789}\mbox{D-i}). By stage 8, the microtubule organization becomes asymmetric, with higher density near the anterior side than at the posterior pole (Fig.~\ref{fig:osk_staufen_stage789}\mbox{D-ii}). This asymmetry persists in stage 9 as the oocyte grows a further twofold (Fig.~\ref{fig:osk_staufen_stage789}\mbox{D-iii}). Over this period, the spatial distribution of \textit{oskar}  mRNA changes significantly: in stage 7, \textit{oskar} is distributed uniformly along the AP axis (Fig.~\ref{fig:osk_staufen_stage789}\mbox{C-ii}, left column), whereas during stages 8 and 9 it progressively shifts toward the posterior pole, where it becomes strongly enriched (Fig.~\ref{fig:osk_staufen_stage789}\mbox{C-ii}, middle and right columns). Staufen protein colocalizes with \textit{oskar} throughout these stages (Fig.~\ref{fig:osk_staufen_stage789}\mbox{C-iii}), making it a suitable proxy for microscopy experiments that cannot be performed directly on \textit{oskar}.

 \begin{figure*}[tbp]\centering\includegraphics[width=1.0\linewidth]{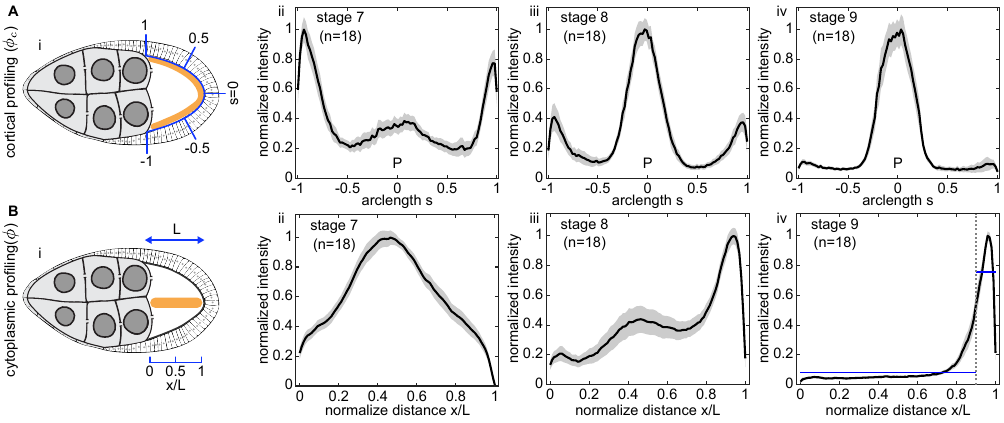}\caption{Redistribution of \textit{oskar} mRNA particles during oocyte development. (A-i) Schematic of the oocyte showing the region for cortical measurement of \textit{oskar} (orange) and the normalized cortical arclength $s$ (blue) measured from the posterior pole (P). (A-ii to A-iv) Normalized intensity of cortical \textit{oskar} as a function of $s$ at stage 7 (A-ii), stage 8 (A-iii), and stage 9 (A-iv). Black lines show the mean and the shaded regions show one standard error of the mean ($n=18$). Normalization is done per stage. (B-i) Schematic of the oocyte showing the region for cytoplasmic measurement of \textit{oskar} (orange) and the normalized distance $x/L$ (blue, $L$ is the oocyte length) measured from the anterior. (B-ii to B-iv) Normalized intensity of \textit{oskar} as a function of $x/L$ at stage 7 (B-ii), stage 8 (B-iii), and stage 9 (B-iv). Black lines show the mean and the shaded regions show one standard error of the mean ($n=18$). Normalization is done per stage. The blue lines in (B-iv) show the spatially averaged intensities in the cytoplasmic bulk ($0.082$; $0 \le x/L \le 0.9$) and in the cortical layer ($0.76$; $0.9 \le x/L \le 1$), separated by the dashed line at $x/L=0.9$.}
\label{fig:osk_profiles_AP_cortex}\end{figure*}

To characterize the dynamics of \textit{oskar} localization from stage 7 to 9, we measured its cortical and bulk distributions in fixed, stained oocytes (see Materials and Methods). For the cortical distribution, we define, along an optical slice, an arclength $-1 \le s \le 1$ with $s=0$ at the posterior pole (Fig.~\ref{fig:osk_profiles_AP_cortex}\mbox{A-i}). At stage 7, we observed mild localization near the posterior pole ($s=0$), with significant enrichment near the anterior corners ($s=\pm1$) (Fig.~\ref{fig:osk_profiles_AP_cortex}\mbox{A-ii}). Through stages 8 and 9, the cortical distribution of \textit{oskar} progressively enriches at the posterior pole and flattens near the anterior side (Fig.~\ref{fig:osk_profiles_AP_cortex}\mbox{A-iii,iv}). To quantify the reorganization of \textit{oskar} in the bulk, we measured the density distribution along the AP axis ($x=0$ corresponds to the anterior side). At stage 7, \textit{oskar} forms a diffuse cloud near the oocyte center with no apparent anterior--posterior bias (Fig.~\ref{fig:osk_staufen_stage789}\mbox{C-ii}, Fig.~\ref{fig:osk_profiles_AP_cortex}\mbox{B-ii}). As the oocyte progresses to stage 8, a clear AP-biased pattern of \textit{oskar} is established (Fig.~\ref{fig:osk_staufen_stage789}\mbox{C-ii}, Fig.~\ref{fig:osk_profiles_AP_cortex}\mbox{B-iii}). By stage 9, the majority of \textit{oskar} mRNA particles localize to the posterior pole (Fig.~\ref{fig:osk_staufen_stage789}\mbox{C-ii}, Fig.~\ref{fig:osk_profiles_AP_cortex}\mbox{B-iv}).

In earlier work, we showed that localization of \textit{oskar}--Staufen RNPs is driven by competition between two cytoskeletal motor proteins: kinesin-1 operating on microtubules, and myosin-V moving along actin filaments \cite{RN24}. Myosin-V retains \textit{oskar}--Staufen RNPs at the cortex, while kinesin-1 transports them away from it. Myosin-V walks along actin filaments, which have no density gradient along the cortex, whereas kinesin-1 walks along microtubules, which show a strong gradient along the AP axis (Fig.~\ref{fig:osk_staufen_stage789}\mbox{C-i},D). Previously, we showed that by perturbing this active cortical exchange, \textit{oskar} tends to aggregate in the bulk cytoplasm \cite{RN24}. For example, in hyperactive kinesin-1 mutant oocytes with enhanced cortical unbinding, the \textit{oskar} RNPs form cytoplasmic droplets. Here, we sought to develop a biophysical model describing the \textit{oskar} cortical localization in the wild type and cytoplasmic aggregation in mutants.

\subsection*{Theory for \textit{oskar} dynamics and localization}

To model the diffusive transport of \textit{oskar} in the cytoplasm, its dynamic binding to and unbinding from the cortex, and its bulk aggregation under perturbations, we propose a two-field continuum model. These two fields interact with one another through boundary conditions that account for the exchange fluxes of \textit{oskar} between the bulk and cortex. We consider the cytoplasm as a binary mixture of \textit{oskar} with concentration $c$ (particle number per unit volume of solution) and molecular volume $\nu$ with cytoplasm (solvent) having an effective molecular volume $\nu_0$. The ratio $\lambda=\nu/\nu_0\sim1100$ is estimated from previously reported super-resolution microscopy of \textit{oskar} RNPs \cite{bose_Cell_2022_liquid-to-solid_phase_transition} (see SI Sec.~3). The local composition of the mixture is therefore characterized by the volume fraction of \textit{oskar} $\phi= c\, \nu$, with  $0 \le \phi \le 1$. The free-energy of the system (\textit{oskar} and solvent) is described by the functional
\begin{align}
    \mathcal{F}(\phi, T) = \int_V \mathrm{d}v \bigg[ f(\phi, T)+\frac{\kappa}{2\,\nu_0}\,|\nabla \phi|^2\bigg]. \label{eq:total_free_energy}
\end{align}
where $V$ is the oocyte volume, $T$ is temperature, $\kappa$ characterizes the contribution to the free-energy due to concentration inhomogeneities, and $f(\phi, T)$ is a binary Flory--Huggins free-energy per unit volume given by
\begin{equation}
    f(\phi, T) = k_B\,T \bigg[ \frac{\phi}{\nu} \ln \phi  + \frac{(1-\phi)}{\nu_0} \ln (1-\phi) +\frac{\chi}{\nu_0} \, \phi(1-\phi)\bigg]. \label{eq:free_energy_flory_huggins}
\end{equation}
with $k_B$ the Boltzmann constant \cite{flory1942thermodynamics, huggins1941solutions, fritschLocalThermodynamicsGovern2021, hymanLiquidLiquidPhaseSeparation2014, julicherDropletPhysicsIntracellular2024}. The first two terms in \eqref{eq:free_energy_flory_huggins} account for the entropic contributions that promote mixing of \textit{oskar} in the solvent. The third term accounts for the enthalpic interactions among \textit{oskar} particles and between \textit{oskar} and the solvent. The interaction parameter $\chi$ promotes demixing when positive. We estimated $\chi\sim0.55$ by fitting the model to microscopy images (see SI Sec.~4), which is a similar value to measurements for P-granule droplets in \textit{C. elegans}  ($\chi=0.505$ at room temperature) \cite{fritschLocalThermodynamicsGovern2021}. The diffusive flux of \textit{oskar} is given by $\bm{J} = -M\boldsymbol{\nabla} \mu$, where $\mu=\nu\,{\delta \mathcal{F}}/{\delta \phi}$ is the exchange chemical potential, and $M=m\,\phi(1-\phi)$ is the mobility describing the molecular exchange of \textit{oskar} with the solvent with a probability proportional to the \textit{oskar} and solvent volume fractions. The bulk evolution of $\phi$ obeys the local conservation law for \textit{oskar}:
\begin{align}
&\partial_t \phi + \boldsymbol{\nabla\,.\,J}  = 0,  &\boldsymbol{x}\in V. \label{eq:conservation_law}
\end{align}
Note that cytoplasmic flows observed in \textit{Drosophila} oocytes during stages 7--9 are relatively weak with no consistent directional pattern \cite{DAHLGAARD2007_DevCel, quinlan2016cytoplasmic, drechsler2017active}. In the study by Khuc Trong \textit{et al.}, they also showed that \textit{oskar} localization is independent of cytoplasmic flows \cite{RN22}. 

We define a cortical layer field $\phi_c$ that represents the volume fraction of surface-bound \textit{oskar} in a cortical layer of thickness $\ell_c$. The \textit{oskar} RNPs bind to the cortex at rate $\beta$ and unbind at rate $\alpha$. Generally, both $\alpha$ and $\beta$ could vary along the cortex. Here, we consider the unbinding rate $\alpha = \alpha(\bm{x})$ to be position-dependent and the binding rate $\beta$ to be constant. 

Because Eq.~(\ref{eq:conservation_law}) is fourth-order in space, it requires two boundary conditions on $\phi$. The first boundary condition is the continuity of RNP flux and the net accumulation of $\phi_c$ on the cortex,
\begin{align}
&\partial_t \phi_c
= \beta\,\phi - \alpha\,\phi_c,& \boldsymbol{x} \in \partial V,\label{eq:phi_s_evolution}\\
&\boldsymbol{n\,.\,J}
= \ell_c\big[\beta\,\phi - \alpha\,\phi_c\big],
& \boldsymbol{x} \in \partial V, \label{eq:BC_cortex_fluxes}
\end{align}
where $\boldsymbol{n}$ is the outward unit normal to the boundary. Note that \eqref{eq:BC_cortex_fluxes} is, through its dependence on $\boldsymbol{J}$, third-order in space and strongly nonlinear. 
The second boundary condition on $\phi$ sets the further interaction of the \textit{oskar}-rich phase with the boundary. We assume a neutral wetting condition between \textit{oskar} and the cortex 
\begin{align}
&\boldsymbol{n \, .\,} \boldsymbol{\nabla}\phi=0,& \boldsymbol{x} \in \partial V,\label{eq:neutral_wetting}
\end{align}
which ensures no preferential interfacial affinity beyond the kinetic exchange described above. This choice of wetting boundary condition does not affect the results we present here. In our simulations, we start from a uniform initial condition $(\phi, \phi_c) = (\phi_0, 0)$ at $t=0$, where $\phi_0$ is constant (see SI Sec.~3). Note that, depending on model parameters, the system has two kinds of steady-state solutions. In one, \textit{oskar} remains mixed in the cytoplasm, while in the other it phase separates and forms a droplet. Below, we provide a phase diagram of these steady states as a function of model parameters.

\subsection*{Control parameters and numerical approach}
There are two length scales and three main time scales that govern the dynamics of the system. The two length scales are the thickness of the RNPs' cortical layer, $\ell_c$, and the cytoplasmic bulk radius, $R$. In stage 9 oocytes, $\ell_c\sim10~\mathrm{\mu m}$ and $R \sim 40~\mathrm{\mu m}$.  The three time scales correspond to the binding time of RNPs to the cortex, $\beta^{-1}$, the unbinding time, $\alpha^{-1}$, and the time it takes for RNPs to diffuse across the oocyte, $t_{d}=R^2/D$, with $D = m\, k_B\, T$ the diffusion coefficient of RNPs. Taking the cytoplasm to be $\sim 1000$ times more viscous than water \cite{doubrovinski2017measurement}, the Stokes--Einstein relation gives a diffusion time of $t_{d} \approx 54~\mathrm{hours}$ for an \textit{oskar} RNP with a radius of order $\sim 100~\mathrm{nm}$. In the next section, we combine photoconversion experiments and theory to estimate the binding and unbinding rates. 

There are six dimensionless groups in the model. The first two are $\chi \sim 0.55$ and $\lambda=\nu/\nu_0 \sim 1100$, which are set by experimental measurements. The remaining four are
\begin{align}
    &A=\alpha\,t_{d}, &&B=\beta\,t_{d}, &\epsilon={\ell_c}/{R},&&\Gamma={\kappa \, m\,\lambda\, t_d}/{R^4}, 
\end{align} 
where $A$ and $B$ represent dimensionless unbinding and binding rates, $\epsilon$ is the ratio of cortical thickness to bulk radius, and $\Gamma$ sets the length scale of gradients in $\phi$. The ratio $r_{\mathrm{ex}} = \alpha/\beta$ is the exchange equilibrium constant that determines the fate of the system by setting the equilibrium ratio of bulk to cortex-bound volume fractions. For large values of $r_{\mathrm{ex}}$ the equilibrium volume fraction of \textit{oskar} in the bulk is larger than that in the cortex. The geometric factor $\epsilon \approx 0.25$ enters in the boundary condition \eqref{eq:BC_cortex_fluxes} and controls the total amount of \textit{oskar} RNPs that the cortex can absorb. The parameter $\Gamma$ controls the critical volume fraction at which the homogeneous mixture is unstable; increasing $\Gamma$ delays the onset of phase separation. 

We solve Eqs.~(\ref{eq:conservation_law}) and (\ref{eq:phi_s_evolution}), subject to boundary conditions (\ref{eq:BC_cortex_fluxes}) and (\ref{eq:neutral_wetting}), using the \texttt{Dedalus} spectral PDE solver in an axisymmetric spherical geometry \cite{Burns_Dedalus_A_flexible_2020} (see SI Sec.~1). Note that, in our model, the two key parameters that determine the equilibrium distribution of \textit{oskar} are the exchange equilibrium constant $r_{\mathrm{ex}}$ and the total amount of \textit{oskar} in the system, which is set by the initial volume fraction $\phi_0$.

\subsection*{Measuring cortical binding and unbinding kinetics}
\begin{figure*}[tbp]
\centering
\includegraphics[width=1.0\linewidth]{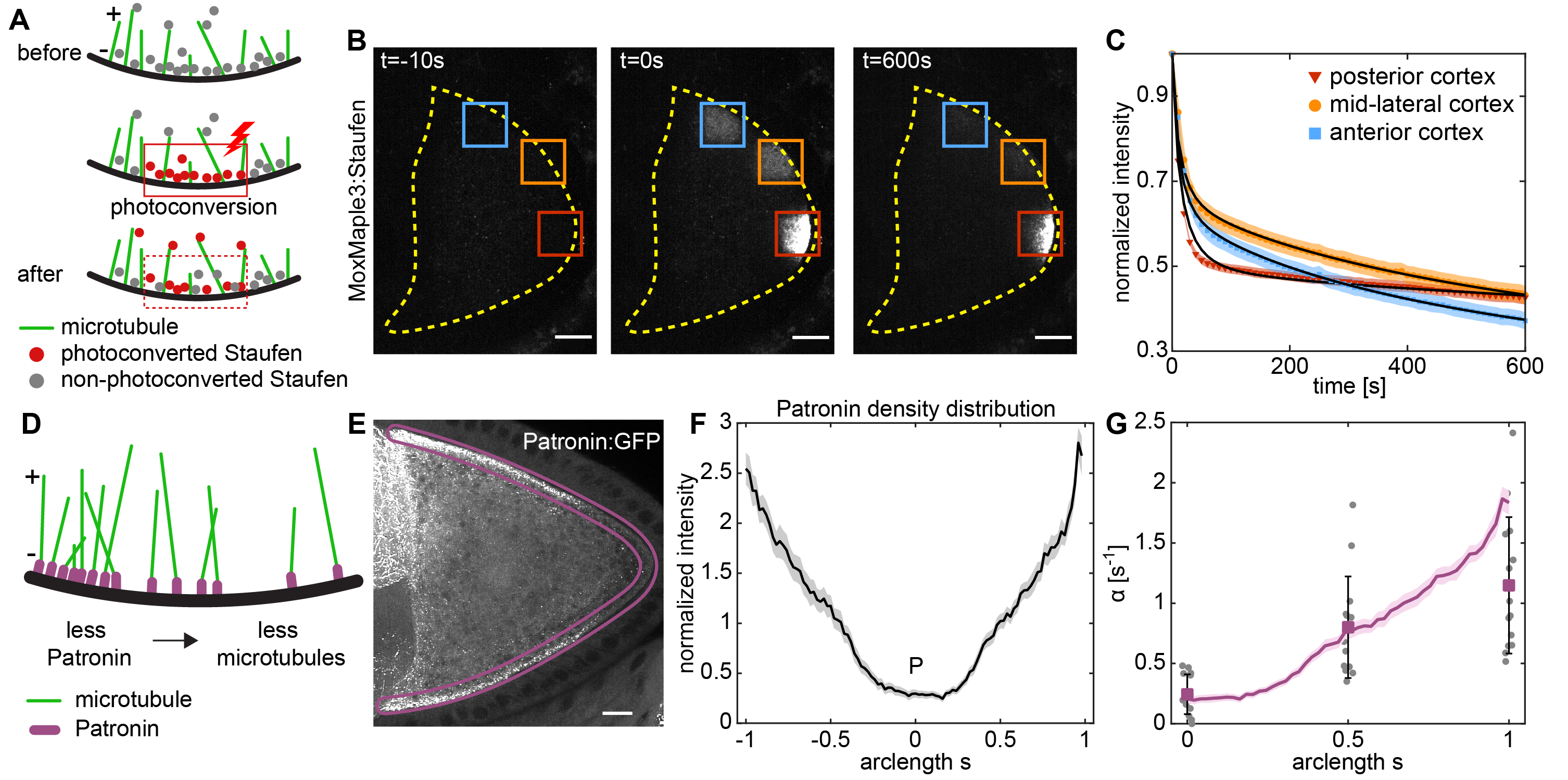}
\caption{Measuring cortical binding and unbinding kinetics of Staufen. (A) Schematic of the photoconversion experiment. The oocyte periphery (black), microtubules (green, with polarity indicated as $\pm$), non-photoconverted Staufen (gray), and photoconverted Staufen (red) are shown. The photoconversion region is indicated as a red box with flash sign. (B) Snapshots from a photoconversion experiment in an oocyte with Staufen-MoxMaple3 taken before photoconversion ($t = -10~\mathrm{s}$), immediately after photoconversion ($t = 0~\mathrm{s}$), and later after photoconversion ($t = 600~\mathrm{s}$) are shown. Three cortical locations (blue, orange, red) of photoconversion and the cell outline (yellow) are indicated. Scale bar is $20~\mathrm{\mu m}$. (C) Average normalized intensity as a function of time at the corresponding photoconversion regions (shaded region, standard error of the mean, $n=14$). The black lines are computed by solving the model with variable unbinding rates between the three photoconverted regions (see SI Sec.~2). (D) Schematic showing the oocyte periphery (black), microtubules (green, with polarity indicated as $\pm$), and Patronin (purple). (E) Patronin-GFP imaged in the mid-plane of a stage 9 oocyte; the cortical region is outlined in purple. Scale bar, $10~\mathrm{\mu m}$. (F) Surface-averaged Patronin intensity profile, $I_p$, as a function of the normalized cortical arclength $s$ measured from the posterior pole (black line, mean; shaded region, standard error of the mean, $n=35$). The posterior pole (P) at $s = 0$ is indicated. (G) Estimation of the unbinding rate profile of \textit{oskar}--Staufen RNPs as a function of normalized arclength $s$. Gray dots show the individual measurements of the unbinding rates from the photoconversion experiments in (B). The average (purple squares) and standard deviation (error bars) are shown. The purple line shows the estimated unbinding rate profile based on Patronin intensity, $\alpha(s) = \alpha_{_\mathrm{SF}}\,I_p(s)$ (solid line, mean; shaded region, standard error of the mean). The scaling factor $\alpha_{_\mathrm{SF}} = 0.67~\mathrm{min^{-1}}$ is obtained by averaging the measurements at $s = 0, 0.5, 1$ (see Sec.~4 in SI).}
\label{fig:unbinding_rate}
\end{figure*}

As established in Fig.~\ref{fig:osk_staufen_stage789}C, Staufen colocalizes with \textit{oskar} and serves as a reliable proxy for tracking its localization. Previously, by artificially shaving cortical microtubules, we showed that Staufen rapidly accumulates at the shaved regions, suggesting that its cortical kinetics are regulated by microtubule density \cite{RN24}. To quantitatively measure these kinetics, we perform a three-point simultaneous photoconversion experiment (Fig.~\ref{fig:unbinding_rate}A). We expressed a photoconvertible Staufen-MoxMaple3 fusion protein in the oocyte; MoxMaple3 switches irreversibly from green to red fluorescence upon exposure to violet light, allowing a spatially defined population of molecules to be labeled and tracked. We performed photoconversion at three distinct cortical regions (anterior, mid-lateral, and posterior cortex) and monitored the decay of red fluorescence at each site (Fig.~\ref{fig:unbinding_rate}B,C). The fluorescence intensity within each region of interest initially decays rapidly, followed by a slower decay at longer times.

We use the model described above to infer Staufen binding and unbinding rates from photoconversion experiments. Given the short duration of the photoconversion experiments, we took a limit of the model considering only diffusion in the bulk and binding and unbinding at the cortex (aggregation in the bulk is neglected). Given the small size of the photoconversion window compared to the oocyte size, we solved the governing equations semi-analytically in a half-space (see SI Sec.~4). We then fit the solution to the experimental decay curves, yielding the measured unbinding rates at the three cortical positions, the cortical binding rate, and the diffusion timescale. We estimated the binding rate as $\beta \sim 2.6~\mathrm{min^{-1}}$ and the unbinding rates at the anterior, mid-lateral, and posterior pole as $\alpha \sim (1.15, 0.80, 0.24)~\mathrm{min^{-1}}$, respectively. The estimated diffusion timescale is $\sim1$ min, corresponding to the time required for an \textit{oskar} RNP to diffuse out of the focal plane (a distance of $\sim1$ $\mathrm{\mu m}$ along the $z$-direction). Using the Stokes-Einstein relation, this yields a particle radius of $100$ nm, in reasonable agreement with the average radius of $200$ nm measured previously by microscopy \cite{bose_Cell_2022_liquid-to-solid_phase_transition}. Model fits at different cortical positions are shown in Fig.~\ref{fig:unbinding_rate}C, and the full list of fit parameters is provided in Table~S2 of the SI.

Previously, we have shown that the cortical unbinding of \textit{oskar}--Staufen RNPs correlates with microtubule density \cite{RN24}. It has been shown that Patronin stabilizes microtubule minus ends and serves as a proxy for cortically anchored microtubule density (Fig.~\ref{fig:unbinding_rate}D) \cite{goodwin2010patronin}. To estimate the unbinding rate profile, we measured the normalized cortical intensity profile of Patronin:GFP, $I_p(s)$, in stage 7 to 9 oocytes (Fig.~\ref{fig:unbinding_rate}E,F, see Fig.~S2 in SI). We observe that in stage 7, Patronin is roughly uniform throughout the cortex, whereas by stage 9 it builds a strong gradient, with $I_p$ dropping by a factor of $\sim9$ from the anterior side to the posterior pole. Note that, in stage 9, the microtubule intensity drops more than twofold along the AP axis (Fig.~\ref{fig:osk_staufen_stage789}\mbox{D-iii}), consistent with Patronin regulating microtubule density. 

We assum that the unbinding rate of \textit{oskar}--Staufen RNPs is proportional to the local Patronin intensity,
\begin{equation}
\alpha(s)=\alpha_{_\mathrm{SF}}\,I_{p}(s), \label{eq:alpha_Ip}
\end{equation}
where $\alpha_{_\mathrm{SF}}$ is a scaling factor. We estimate $\alpha_{_\mathrm{SF}}$ from photoconversion experiments and Patronin measurements as follows. Photoconversion provides the unbinding rate $\alpha$ at three cortical positions, $s = 0, 0.5, 1$. Combining these with the corresponding Patronin intensities $I_p(s = 0, 0.5, 1)$ gives an average value of $\alpha_{_\mathrm{SF}} \sim 0.67~\mathrm{min^{-1}}$. The resulting unbinding rate profile $\alpha(s)$ across the cortex is shown in Fig.~\ref{fig:unbinding_rate}G.

\begin{figure*}[tbp]%
\centering
\includegraphics[width=1\linewidth]{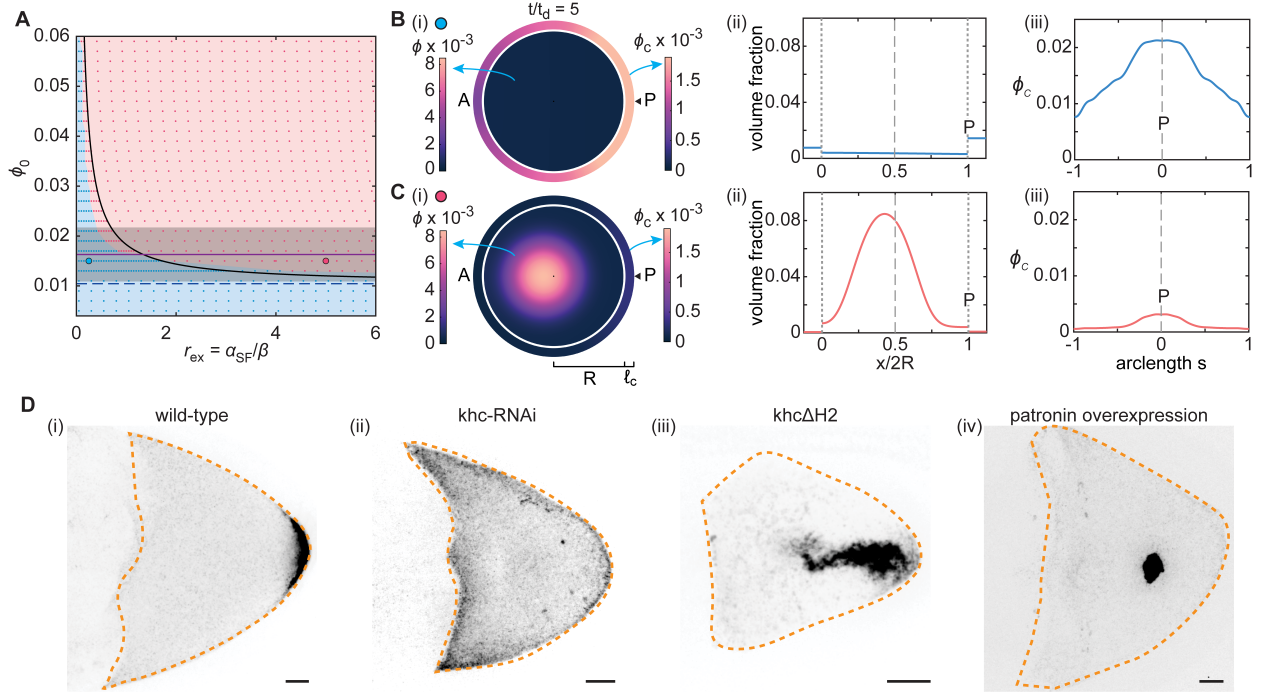}
\caption{Phase diagram of the \textit{oskar} localization pattern and model predictions for the wild-type and mutants. (A) Phase diagram of the system as a function of $\phi_0$ and $r_{\mathrm{ex}}$, using parameters estimated from experiments (SI Secs.~4--6, Table~S2). The dots show individual simulations. The purple line and the gray shaded region indicate the mean volume fraction $\phi_0 = 0.016 \pm 0.0054$ and one standard deviation, estimated from experiments (SI Sec.~5). The dashed blue line is the spinodal volume fraction for a system without cortical binding and unbinding, and the solid black line is the spinodal volume fraction for a system with uniform cortical binding and unbinding. Two circles denote simulations with $(\phi_0,r_{\mathrm{ex}}) = (0.015, 0.25)$ (blue) and $(\phi_0,r_{\mathrm{ex}}) = (0.015, 5)$ (red). (B,C) Snapshots at $t/t_d=10$ of simulations for the two systems marked in (A). The $\phi$ field in a cross-section of the cytoplasm and $\phi_c$ in the cortical layer (i). The \textit{oskar} profile along the AP axis, with the cortical layer (thickness of $\epsilon/2=0.125$) marked by two dotted lines at $x/2R=0, 1$ (ii). The cortical profile of \textit{oskar} as a function of arclength $s$ measured from the posterior pole (iii).  Time is nondimensionalized by the diffusion time across the oocyte and the posterior pole is marked by P. (D) \textit{oskar} mRNA localization patterns in wild-type (i), kinesin-1 heavy chain (Khc) knockdown (ii), hyperactive kinesin-1 (khc$\mathrm{\Delta H2}$) mutant (iii), and Patronin-overexpressing (iv) oocytes are shown. Orange lines indicate the oocyte periphery. Scale bars, $10~\mathrm{\mu m}$.}
\label{fig:phase_diagram}
\end{figure*}

\subsection*{Localization patterns in the wild-type and mutants}
Figure~\ref{fig:phase_diagram}A presents the phase diagram of the steady-state model system, which exhibits two distinct regimes. In the first (blue), \textit{oskar} accumulates at the posterior cortex while remaining homogeneously distributed in the bulk. In the second (red), the bulk volume fraction is high enough for \textit{oskar} to phase separate into a cytoplasmic droplet. The system transitions between these two regimes when the bulk volume fraction exceeds the spinodal volume fraction, beyond which the free-energy of the mixture has negative curvature ($\partial^2 f/\partial\phi^2 < 0$) and the homogeneous state is linearly unstable. The dashed line marks the spinodal of an isolated system with no cortical exchange ($A=B=0$), $\phi_{_\mathrm{sp1,sp2}}= [b\pm(b^2-8\chi/\lambda)^{1/2}]/4\chi, ~b=2\chi+\lambda^{-1}-1$. The solid black line denotes the shifted spinodal volume fraction for a system with uniform cortical binding and unbinding, $\phi^* = (1+3\epsilon \beta/\alpha)\, \phi_{_\mathrm{sp1}}$. The purple line and the shaded region in Fig.~\ref{fig:phase_diagram}A highlight the total volume fraction of \textit{oskar}--Staufen RNPs estimated from stage 9 oocytes ($\phi_0 = 0.016 \pm 0.0054$, mean $\pm$ std) (see SI). 

In Fig.~\ref{fig:phase_diagram}B, we show a simulation with the total amount of \textit{oskar} similar to the experimental measurements ($\phi_0=0.015$) with an exchange equilibrium constant estimated from photoconversion experiments ($r_{\mathrm{ex}}=0.25$, blue circle in Fig.~\ref{fig:phase_diagram}A). The simulation starts from a uniform \textit{oskar} distribution in the bulk at $t=0$. The corresponding $\phi$ and $\phi_c$ fields at $t/t_d=10$ are shown in Fig.~\ref{fig:phase_diagram}\mbox{B-i}. We observe that \textit{oskar} localizes at the posterior cortex with a relatively uniform distribution in the bulk, which is similar to the localization pattern in wild-type oocytes (Fig.~\ref{fig:phase_diagram}\mbox{B-ii,iii,D-i}, and Fig.~\ref{fig:osk_profiles_AP_cortex}\mbox{A-iv,B-iv}). With the total amount of \textit{oskar} in the system kept at the same level, in the limit where the unbinding rate is zero ($r_{\mathrm{ex}}=0$), all of the \textit{oskar} is eventually depleted from the bulk and accumulates along the cortex. Because the binding rate is spatially uniform, in this regime, we expect no posterior bias. To test this, we knocked down the kinesin-1 motor and observed a uniform enrichment of \textit{oskar} at the cortex (Fig.~\ref{fig:phase_diagram}\mbox{D-ii}). This is consistent with the prediction of the model.  

We next perform a simulation where we keep the total amount of \textit{oskar} the same as before ($\phi_0=0.015$), but increase the exchange equilibrium constant ($r_{\mathrm{ex}}=5$, red circle in Fig.~\ref{fig:phase_diagram}A). In this simulation, the \textit{oskar} RNPs have shorter residence times on the cortex. As a result, \textit{oskar} is released from the cortex into the cytoplasm, raising the average bulk volume fraction, causing \textit{oskar} to phase separate and form a droplet in the cytoplasm (Fig.~\ref{fig:phase_diagram}\mbox{C-i,ii}).  Under this condition, the amount of \textit{oskar} in the cortical layer is significantly lower than the wild-type (Fig.~\ref{fig:phase_diagram}\mbox{B-iii}, C-iii). 

To test this prediction, we performed two genetic perturbation experiments. In the first, we increased kinesin-1 activity directly by using a constitutively active mutant, Khc$\Delta$H2 (referred to here as \textit{hyperactive kinesin-1}). This mutant kinesin-1 lacks the flexible hinge region and therefore cannot adopt its auto-inhibited conformation, which causes it to be always active (see Materials and Methods)~\cite{BARLAN2013CurrentBiology, RN25, RN24}. Experiments in cultured mammalian cells show that only $\sim30\%$ of kinesin-1 motors are active and exert force, while the rest are in their auto-inhibited state and do not exert force \cite{kinesin_abundance_JCB1989}. In addition, it has been shown that this hyperactive kinesin-1 motor moves $\sim2.5$ times faster than the wild-type ($20.8\pm4.5~\mu \mathrm{m/min}$ versus $8.1\pm3.9~ \mu\mathrm{m/min}$)~\cite{friedman1999single}. We therefore estimated that the effective unbinding rate in the hyperactive kinesin-1 mutant is roughly an order-of-magnitude higher than the wild-type. We observe that in this mutant, \textit{oskar} phase separates and forms a droplet in the cytoplasm (Fig.~\ref{fig:phase_diagram}\mbox{D-iii}). This is consistent with the prediction of the model (Fig.~\ref{fig:phase_diagram}C).  

We further examined the effect of active cortical exchange on \textit{oskar} localization by overexpressing Patronin, which stabilizes microtubule minus ends against depolymerization by Klp10A. The resulting increase in microtubule stability provides more persistent tracks for kinesin-1 and thus removing \textit{oskar} more rapidly from the cortex. In this mutant, we also observe that \textit{oskar} phase separates and forms a droplet in the cytoplasm (Fig.~\ref{fig:phase_diagram}\mbox{D-iv}), which is in agreement with the model prediction. Note that this cytoplasmic phase separation caused by increasing the unbinding rate (and therefore increasing $r_{\mathrm{ex}}$) only occurs when the total amount of \textit{oskar} is beyond the spinodal threshold, $\phi_0> \phi_{\mathrm{sp1}}$ (dashed line in Fig.~\ref{fig:phase_diagram}A). 

\subsection*{Effect of asymmetric oocyte growth on \textit{oskar} localization}
In all of our experiments with hyperactive kinesin-1 and Patronin overexpression, the \textit{oskar} droplet is positioned closer to the posterior pole, with none touching the anterior side of the oocyte. We examined the long-time (say, $t/t_d > 10$) positioning of phase-separated droplets in our simulations and observed that the droplet drifts toward the anterior pole, eventually reaching and wetting the anterior cortex. This slow drift arises from the anterior--posterior asymmetry in cortical unbinding: on the posterior side of the droplet, where the unbinding rate is lower, cytoplasmic \textit{oskar} is taken up by the cortex for longer than on the anterior side. The asymmetric depletion shifts the droplet's mass balance and displaces it away from the posterior pole. This anterior drift is inconsistent with experimental observations.  Moreover, this drift occurs at timescales that are biologically irrelevant ($t\sim 540$ hours versus the oocyte lifetime of $\sim 12$ hours from stage 7 to 9) and therefore negligible for droplet repositioning during oogenesis.  

\begin{figure}[tbp]
\centering
\includegraphics[width=0.65\linewidth]{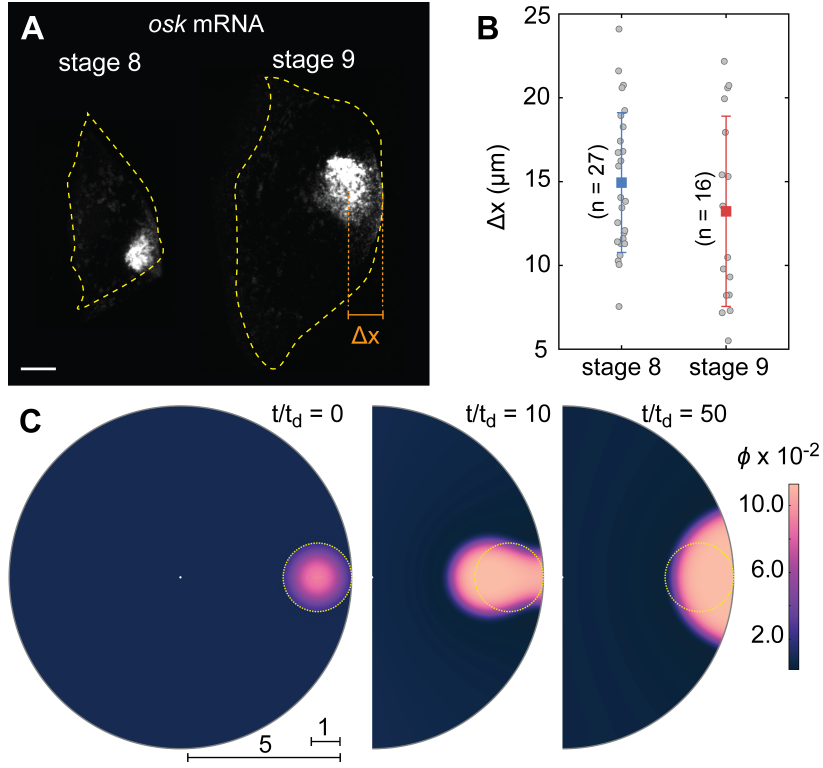}
\caption{Posterior retention of the \textit{oskar} droplet under asymmetric oocyte growth. (A) Phase-separated \textit{oskar} droplet at stages 8 and 9 in the hyperactive kinesin-1 mutant. Scale bar, $10~\mu\mathrm{m}$. The distance between the droplet centroid and the posterior pole along the anterior--posterior axis ($\Delta x$) and the oocyte outline (yellow line) are indicated. (B) Individual measurements of $\Delta x$ for stage 8 and 9 oocytes (gray dots), and the average (squares) and standard deviations (error bars) are shown. (C) Snapshots of the $\phi$ field at $t/t_d=0, 10, 50$ from a simulation with $r_{\mathrm{ex}}=10$ in a sphere of radius $R_f = 5$. The initial condition $\phi(t=0)$ is taken from the equilibrium $\phi$ field in a spherical domain of radius $R_i = 1$ (outlined with yellow dashed lines) under a uniform unbinding rate (see SI Sec.~7). The diffusion time scale $t_d$ is the time required for \textit{oskar} particles to travel a distance of the oocyte radius. }
\label{fig:drop_positioning}
\end{figure}

To characterize the positioning of the \textit{oskar} droplet, we measured the distance between the droplet centroid and the posterior pole in $n=27$ stage 8 and $n=16$ stage 9 oocytes with hyperactive kinesin-1 (Fig.~\ref{fig:drop_positioning}A,B). We observe that the droplet is located near the posterior pole and occasionally wets the cortex (Fig.~\ref{fig:phase_diagram}G, Fig.~\ref{fig:drop_positioning}A). Interestingly, the distance between the droplet centroid and the posterior pole shows no significant difference between stage 8 and stage 9, even though the oocyte grows during this period (Fig.~\ref{fig:drop_positioning}B). If the oocyte growth were uniform in all directions, we would expect the distance between the droplet and the posterior pole to increase, which we did not observe in experiments. We therefore infer the oocyte growth to be asymmetric, such that the anterior side, where the unbinding rate is higher, moves away from the \textit{oskar} droplet. In contrast to the simulation in a fixed domain, where we observe the anterior drift of the droplet, here we expect the droplet to position differently.  

To capture the effect of this asymmetric growth, we performed simulations in two spherical domains of dimensionless radii $R_i = 1$ and $R_f = 5$ (similar to the change in oocyte size between stages 7 and 9; Fig.~\ref{fig:osk_staufen_stage789}B). The smaller domain represents the oocyte at stage 7, when it is small and lacks a microtubule gradient, so that cortical binding and unbinding are uniform. The larger domain represents the oocyte at stage 9, after growth, when the established microtubule gradient imposes the biased unbinding rate of \eqref{eq:alpha_Ip}. We seed the larger domain with the equilibrium solution obtained in the smaller one, placing it at the posterior pole (Fig.~\ref{fig:drop_positioning}C; see SI). As the simulation proceeds, the phase-separated droplet grows while drifting toward the anterior. The competition between this anterior drift and the droplet's growth sets the final position and determines whether the droplet reaches and wets the posterior pole (Fig.~\ref{fig:drop_positioning}C). Note that in our simulations, by varying $\phi_0$, we observe cases where the droplet does not wet the posterior pole, consistent with the occasional cases we observed in the hyperactive kinesin-1 mutant. 

\section*{Discussion}

Our results provide a quantitative framework for understanding how competition between kinesin-1 and myosin-V generates posterior localization of \textit{oskar} mRNA in the \textit{Drosophila} oocyte. Myosin-V promotes recruitment of \textit{oskar}--Staufen RNPs to the actin-rich cortex, whereas kinesin-1 removes them from cortically anchored microtubules and releases them into the bulk. Perturbing this balance, as in the hyperactive kinesin-1 and Patronin-overexpression mutants, drives phase separation of RNPs in the bulk. The description of membraneless organelles and protein aggregates through the equilibrium thermodynamics of liquid--liquid phase separation has substantially advanced our understanding of how cells compartmentalize their interiors \cite{brangwynne2009germline, fritschLocalThermodynamicsGovern2021, alberti2019considerations}. Our model yields an estimate of the enthalpic interaction of \textit{oskar} RNPs, $\chi\approx0.55$, close to the value $\chi\approx0.505$ obtained previously for P granules in \textit{Caenorhabditis elegans} using temperature perturbations. This agreement suggests that the two systems may be governed by comparable underlying energetics. 

Patterning in early embryos is a complex process in which several mechanisms, such as binding and unbinding, diffusion, and directed transport along cytoskeletal filaments, operate simultaneously. Isolating the effects of these mechanisms and measuring their biophysical parameters inside a living cell are inherently challenging. Several theoretical studies have aimed to understand the physical basis of patterning in cells \cite{hawkins2009rebuilding, bressloff2015stochastic, ciocanel2018modeling}. However, a quantitative understanding of patterning in embryos requires a direct feedback between theory and experiments. The binding and unbinding rates and diffusion coefficient estimated here are one such example, as the contribution of any single process cannot easily be isolated from the others. Combining theory and experiment offers a way to disentangle these processes, as shown previously for PAR domain formation in \textit{C. elegans}, where cortical flows advect PAR proteins while a reaction--diffusion exchange couples the bulk and the cortex \cite{goehring2011polarization}. Here, we take a similar approach, combining theory and photoconversion experiments to estimate the cortical binding rate of \textit{oskar} RNPs as $\beta\approx 2.6~\mathrm{min}^{-1}$ and the unbinding rates at three cortical regions, which vary by a factor of five between the anterior and posterior regions ($\alpha_{\mathrm{a}}\approx1.15~\mathrm{min}^{-1}$, $\alpha_{\mathrm{p}}\approx0.24~\mathrm{min}^{-1}$). At the posterior, the binding rate is therefore roughly 20-fold faster than the unbinding rate. We also measured the passive diffusion coefficient of \textit{oskar}--Staufen RNPs to be $\sim1~\mathrm{min}^{-1}$, which allowed us to estimate their size to be $\sim100~\mathrm{nm}$. This set of measurements constrains the model entirely from experiment.  

Presumably, robust formation of germ cells requires a certain amount of \textit{oskar} mRNA in the posterior cap. Our model shows that this requirement is not free of cost: above a threshold amount of \textit{oskar}, set by the binding and unbinding rates, additional \textit{oskar} RNPs are no longer captured by the cortex and instead phase separate into a droplet in the bulk, competing with the very cap they are meant to build. The system's behavior is therefore governed by two key parameters: the total amount of \textit{oskar} in the oocyte, set by $\phi_0$, and the exchange equilibrium constant $r_{\mathrm{ex}}$. For a given exchange equilibrium constant, increasing $\phi_0$ drives liquid-liquid phase separation of \textit{oskar} RNPs and the formation of a droplet in the bulk, with weakened localization at the cortex. At the physiological value of $\phi_0$ estimated here ($\phi_0 = 0.016 \pm 0.005$), the exchange equilibrium constant $r_{\mathrm{ex}}$ can vary by roughly eightfold without driving \textit{oskar} into the phase-separated regime, indicating that the localization pattern is robust to variation in cortical exchange kinetics at this \textit{oskar} dose. The phase diagram shows, however, that this robustness is itself dose-dependent: as $\phi_0$ increases, the range of $r_{\mathrm{ex}}$ compatible with a purely cortical cap narrows substantially, and an increase in $\phi_0$ by a factor of two is sufficient to trigger phase separation in the bulk. This asymmetry raises the possibility that the amount of \textit{oskar} mRNA produced by the nurse cells and delivered to the oocyte is under selective constraint, tuned not only to supply a sufficient posterior cap, but also to keep the system within the low-$\phi_0$ regime where its localization pattern remains robust to natural variation in motor activity.

By stage 10 of oogenesis, the oocyte cytoplasm is no longer stationary: cell-spanning flows arise from motors carrying cargo along cortically anchored microtubules  \cite{gutzeit1982time, RN7}. Through a combination of modeling, large-scale simulation, and live imaging, we now have a comprehensive, quantitative understanding of how these hydrodynamically organized flows, termed twisters, are generated \cite{monteith2016mechanism,stein2021swirling,dutta2024self,chakrabarti2024cytoplasmic,jain2025geometric}.  What function might these flows serve for \textit{oskar} localization? Recently, we showed that these cytoplasmic flows strongly facilitate mixing and significantly speed up delivery of molecules and particles across the oocyte \cite{jain2026travelingwavesenhancetransport}: at characteristic speeds of 100--300 nm/s, a twister could carry \textit{oskar} RNPs across the oocyte in minutes, compared with tens of hours required by diffusion alone. This could help to maintain the cap as the oocyte further grows and the nurse cells dump their contents into the oocyte. A second possibility is suggested by genetic perturbations such as \textit{Klc-RNAi} (kinesin-1 light chain) and \textit{btz-RNAi} (Barentsz), which cause Staufen to become mislocalized in the cytoplasm at stage 9 without significantly affecting cytoplasmic streaming \cite{RN8}. In these mutants, streaming partially resolves the early mislocalization, redistributing Staufen back into a more coherent posterior cap by stage 10. Thus, these twister flows can act as a correcting mechanism, redistributing mislocalized or aggregated \textit{oskar}--Staufen RNPs back into a coherent posterior cap. An extension of the model to include these cortically generated flows, with \textit{oskar} binding, unbinding, and phase separation in the presence of twister flows, is a compelling direction for future work.

\section*{Materials and Methods}

\subsection*{Induction of Germline Clone of Khc$\mathrm{\Delta}$H2} \label{sec:hyperactive_kinesin}
The Khc[$\mathrm{\Delta}$H2] fly line was a generous gift from Dr. Jill C. Wildonger (University of California, San Diego) \cite{RN25}. To generate Khc$\mathrm{\Delta}$H2 germline clones, virgin females of genotype FRTG13 Khc$\mathrm{\Delta}$Hinge2/CyO were crossed with males carrying hs-flp[12]/Y; FRTG13 ubi-GFP.nls/CyO. Progeny pupae were heat-shocked at $37^\circ\mathrm{C}$ for 2 hours on day 7 or 8 after egg laying (AEL) to induce mitotic recombination. Following heat shock, non-CyO F$_1$ females were collected 3--4 days later, fed with dry active yeast overnight to promote ovarian development, and dissected for examination. Khc$\mathrm{\Delta}$H2 germline clones were identified by the absence of nuclear GFP signal in the germline cells of the egg chamber.
\subsection*{Single-Molecule Inexpensive Fluorescence in situ Hybridization (smiFISH)}
Single-molecule inexpensive FISH (smiFISH) was performed following established protocols \cite{RN26,RN27}. A set of twenty-five 20-nucleotide DNA probes complementary to the \textit{oskar} coding region and 3$^{\prime}$ UTR was designed, each appended with a 3$^{\prime}$ FLAP-X sequence (5$^{\prime}$-CCTCCTAAGTTTCGAGCTGGACTCAGTG-3$^{\prime}$). The unlabeled \textit{oskar}-specific probes were mixed at equimolar ratios and annealed to a fluorescently labeled FLAP-X probe carrying Cy5 modifications at both the 5$^{\prime}$ and 3$^{\prime}$ ends (/5Cy5/CACTGAGTCCAGCTCGAAACTTAGGAGG-/3Cy5Sp/).

\subsection*{Photoconversion of Staufen-MoxMaple3} \label{sec:photoconvertible_staufen}
The pUASp-Staufen-MoxMaple3 construct was generated by inserting Staufen and MoxMaple3 into the pUASp vector via KpnI(5’)/NotI(3’), and NotI(5’)/XbaI(3’), respectively. UASp-Staufen-MoxMaple3 transgenic lines were generated by random P-element insertion by BestGene, Inc. To express Staufen-MoxMaple3 in the oocyte, males carrying UASp-Staufen-MoxMaple3 were crossed with virgin females of the germline-specific driver line nos-Gal4[VP16] at room temperature. $\mathrm{F_1}$ females were collected, fed with dry active yeast overnight to promote oogenesis, and used for live imaging. Ovaries were dissected and mounted in Halocarbon oil 700 as previously described \cite{RN26,RN28}. Live imaging was performed on a Nikon W1 spinning-disk confocal microscope (Yokogawa CSU, pinhole size $50~\mathrm{\mu m}$) equipped with a Hamamatsu ORCA-Fusion digital CMOS camera and a $40\times/1.25~\mathrm{NA}$ silicone-oil immersion objective. MoxMaple3 fluorescence was photoconverted from green to red at three cortical sites of each oocyte (anterior, mid-lateral, and posterior cortex) using $405~\mathrm{nm}$ light patterned by a Mightex Polygon DMD illuminator. After photoconversion, red fluorescence signals of Staufen-MoxMaple3 were recorded every 10 seconds for a total of 10 minutes. Fluorescence intensities at all three cortical regions were quantified using Fiji/ImageJ.

\subsection*{\textit{Drosophila} Husbandry}
Fly stocks and crosses were maintained on standard cornmeal food (Nutri-Fly Bloomington Formulation, Genesee Scientific, Cat. \#66-121) supplemented with active dry yeast at room temperature ($\sim22^\circ\mathrm{C}$). The following fly stocks were used in this study: $\mathrm{w}^{[1118]}$ (X, Bloomington Drosophila Stock Center/BDSC \#5905); $\mathrm{nos-Gal4}^{[\mathrm{VP16}]}$ (III) \cite{van1998regulation,lu2012niche}; UASp-InfraRed-$\alpha$tub (III) and Staufen-mCherry (CRISPR knock-in line), both provided by Dr. R\'egis Giet (Institute of Genetics and Development of Rennes, France); UASp-GFP-Patronin (II) \cite{RN24,RN12,RN28}; UAS-Khc-RNAi (attP2, III, TRiP.GL00330, BDSC \#35409, targeting nucleotides 150--170 of the Khc coding sequence) \cite{RN7}; mat $\alpha$tubGal4-VP16[V2H] (II, BDSC \#7062); hs-FLP[12] (X, BDSC \#1929); FRTG13 ubi-GFP.nls (II, BDSC \#5826); FRTG13 Khc$\mathrm{\Delta}$H2 (II) \cite{RN24}. The UASp-Staufen-MoxMaple3 (II) line was generated by BestGene Inc. using standard P-element--mediated transformation.

\section*{Acknowledgments}
We thank Dr. Régis GIET (Institute Genetics \& Development of Rennes, France) and the Bloomington Drosophila Stock Center (supported by NIH P40OD018537) for fly stocks. We thank all current and past members of the Gelfand laboratory for their support, discussion, and suggestions. This work was supported by the National Institute of General Medical Sciences (2R35GM131752 to V.I.G.). We acknowledge support from the CCBx program of the Center for Computational Biology of the Flatiron Institute. The computations in this work were performed at facilities supported by the Scientific Computing Core at the Flatiron Institute, a division of the Simons Foundation.

\section*{Author contributions}
M.J.S and V.I.G designed the project. W.L. and M.L. performed the experiments. M.F. and M.J.S developed the model and performed the simulations. M.F, W.L., and R.F. analyzed the data and prepared the figures. M.F., R.F., W.L., and M.J.S wrote the first draft. All authors contributed to the final draft.

\section*{Competing interests}
The authors declare no competing interest.

\bibliographystyle{unsrt-keepcase}
\bibliography{Refs}

\end{document}


\title{Supporting Information for:
Polarization of the \textit{Drosophila} oocyte by active cortical exchange}

\author[a]{Mohammadhossein Firouznia}
\author[b]{Wen Lu}
\author[b]{Margot Lakonishok}
\author[a]{Reza Farhadifar}
\author[b]{Vladimir~I.~Gelfand}
\author[a,c]{Michael~J.~Shelley}

\affil[a]{Center for Computational Biology, Flatiron Institute, New York, NY 10010}
\affil[b]{Department of Cell and Developmental Biology, Feinberg School of Medicine, Northwestern University, Chicago, IL 60611}
\affil[c]{Courant Institute, New York University, New York, NY 10012}

\date{}

\maketitle

\section{Anterior-to-posterior distance from stage 7 to stage 9}
We quantified the changes in oocyte size by fixing $n = 18$ oocytes at stages 7--9 and staining them with DAPI. We measured the oocyte length as the distance from the anterior side (shared with the nurse cells) to the posterior pole in the mid-plane of the oocyte.
\begin{figure*}[h]
\centering
\includegraphics[width=0.4\linewidth]{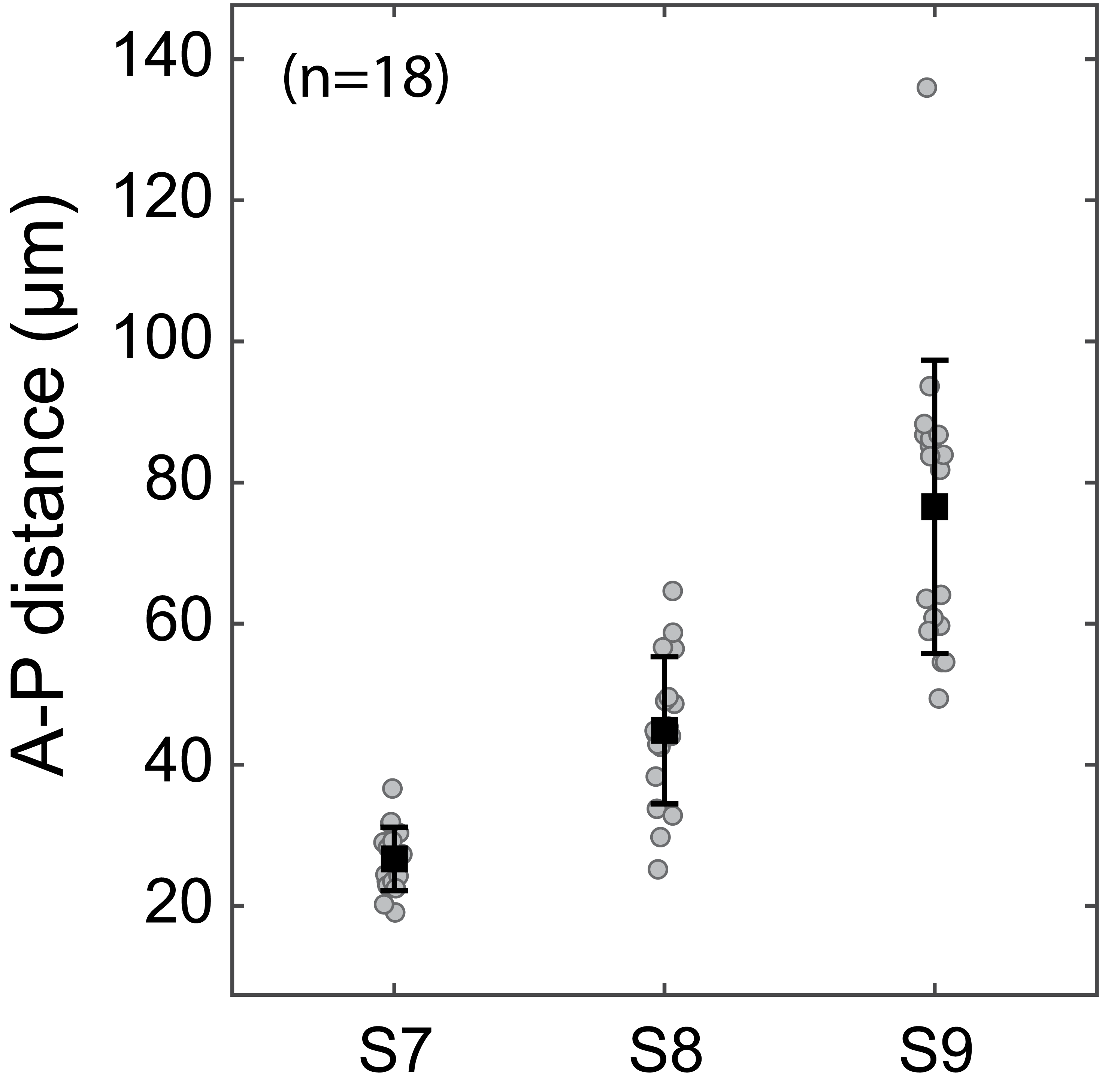}
\caption{Length of A-P axis of  oocytes at stage 7  ($27 \pm 5~\mathrm{\mu m}$, $n=18$), stage 8 ($45 \pm 10~\mathrm{\mu m}$, $n=18$), and stage 9 ($77 \pm 21~\mathrm{\mu m}$, $n=18$). Values are mean $\pm$ one standard deviation.}
\label{fig:A-P_distance_stage789}
\end{figure*}
\FloatBarrier
\pagebreak

\section{Cortical Patronin profiles from stage 7 to stage 9}
We measured the normalized cortical intensity profile of Patronin-GFP, $I_p(s)$, in stage 7 to 9 oocytes, where the arclength $-1 \le s \le 1$ is measured from the posterior pole ($s=0$).
\begin{figure*}[h]
\centering
\includegraphics[width=1.0\linewidth]{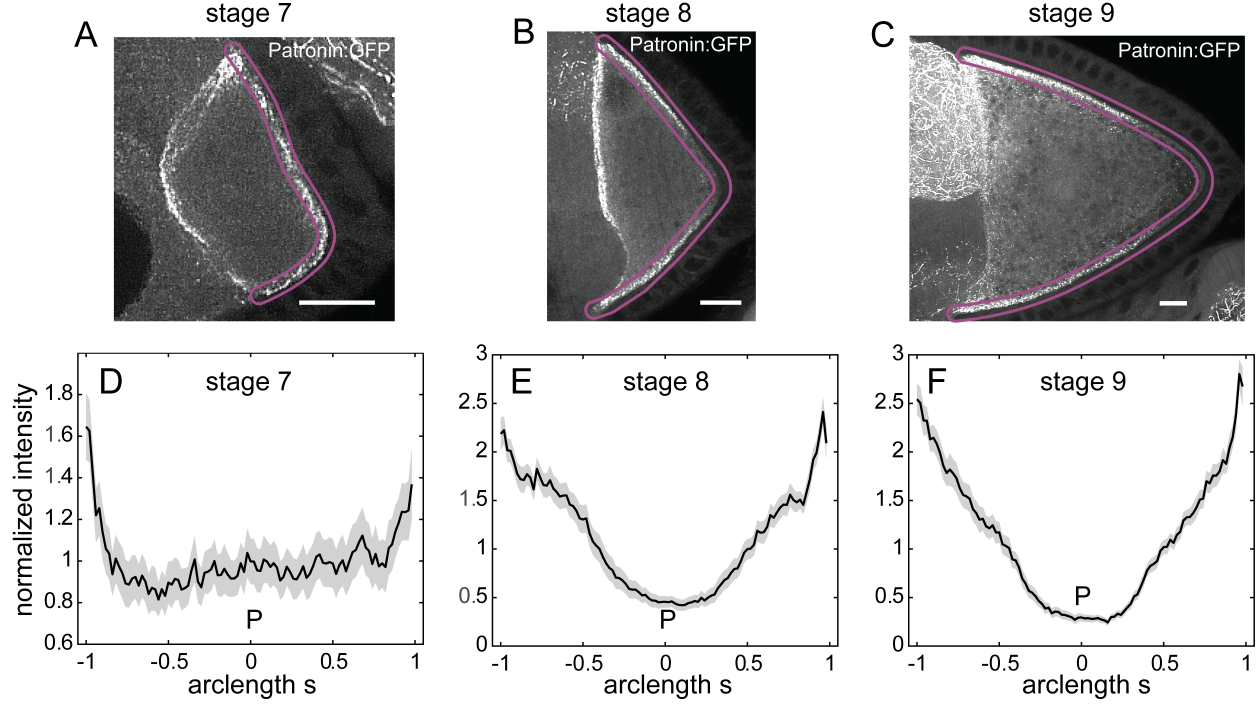}
\caption{(A--C) Representative images of Patronin-GFP in the mid-plane of stage 7 (A), stage 8 (B), and stage 9 (C) oocytes; the cortical region is outlined in purple. Scale bar, $10~\mathrm{\mu m}$. (D--F) Surface-averaged Patronin intensity profile, $I_p$, as a function of the normalized cortical arclength $s$ measured from the posterior pole, at stage 7 (D; $n=35$), stage 8 (E; $n=37$), and stage 9 (F; $n=35$) (black line, mean; shaded region, standard error of the mean). The profile is normalized so that its average over the cortical surface is unity, $\int_{\partial V}\!\! I_p \, \mathrm{d}A \, / \int_{\partial V} \!\!\mathrm{d}A = 1$. The posterior pole (P) at $s = 0$ is indicated. }
\label{fig:patronin_stage789}
\end{figure*}
\FloatBarrier

\pagebreak

\section{Model for Transport and Cortical Exchange of \textit{oskar} mRNA}

As discussed in the main text, we propose a continuum model for the cytoplasmic transport and cortical exchange of \textit{oskar} during stages 7--9 of \textit{Drosophila} oogenesis based on local thermodynamic equilibrium. The model describes two coupled processes: (i) phase separation of \textit{oskar} and (ii) its reversible association with the cortex through binding and unbinding.  We consider a binary mixture of \textit{oskar} with concentration $c$ and molecular volume $\nu$ in a solvent with an effective molecular volume $\nu_0$ that represents the rest of the cytoplasmic fluid. The local composition of the mixture is therefore characterized by volume fraction $\phi= c\, \nu$, with  $0 \le \phi \le 1$. The phase separation of \textit{oskar} in the cytoplasm can be described by an effective free energy which is a function of local volume fraction $\phi$ and temperature $T$:
\begin{align}
    \mathcal{F}(\phi, T) = \int_V \mathrm{d}v \bigg[ f(\phi, T)+\frac{\kappa}{2\,\nu_0} |\nabla \phi|^2\bigg]. \label{eq:total_free_energy}
\end{align}
Here, $\kappa$ characterizes the change in free energy due to concentration inhomogeneities, and $f(\phi, T)$ is a binary Flory--Huggins free energy per unit volume given by 
 \cite{flory1942thermodynamics, huggins1941solutions, fritschLocalThermodynamicsGovern2021, hymanLiquidLiquidPhaseSeparation2014, julicherDropletPhysicsIntracellular2024}:
\begin{equation}
    f(\phi, T) = k_B\,T \bigg[ \frac{\phi}{\nu} \ln \phi  + \frac{(1-\phi)}{\nu_0} \ln (1-\phi) +\frac{\chi}{\nu_0} \, \phi(1-\phi)\bigg]. \label{eq:free_energy_flory_huggins}
\end{equation}
The effective interactions among \textit{oskar} molecules and between \textit{oskar} and the solvent are characterized by $\chi(T)$, with $\chi>0$ promoting demixing. In our model, we consider a fixed spherical domain with radius $R$. The diffusive flux of \textit{oskar} is given by $\bm{J} = -M\bnabla \mu$, where $\mu=\nu\,{\delta \mathcal{F}}/{\delta \phi}$ is the exchange chemical potential, and $M=m\,\phi(1-\phi)$ is the mobility describing the molecular exchange of \textit{oskar} with the solvent with a probability proportional to \textit{oskar} and solvent volume fractions. The evolution of $\phi$ obeys the conservation of fluxes of \textit{oskar}:
\begin{align}
&\partial_t \phi + \boldsymbol{\nabla\,.\,} \boldsymbol{J} = 0,&\bm{x}\in V.
\end{align}
Defining $D=m\,k_B T$, $\gamma=\kappa \,m \, \lambda$, $\lambda={\nu}/{\nu_0}$, the transport equation becomes:
\begin{align}
&\partial_t \phi  =D\, \Delta \big[ \phi +  g(\phi) \big]-\gamma\, \bnabla \bcdot \big[\phi(1-\phi) \, \bnabla(\Delta \phi) \big], &\bm{x}\in  V,\label{eq:transport_cf}\\
&g(\phi)=\Big[ (\lambda-2\,\lambda\, \chi-1) \dfrac{\phi^2}{2} +2\lambda\, \chi\, \dfrac{\phi^3}{3} \Big].
\end{align}
The parameters $\lambda$ and $\chi$ are inferred from experimental measurements, as described in Secs.~\ref{sec:SI:total_osk} and \ref{sec:SI:fitting_of_thermodynamic_model}. The boundary conditions are:
\begin{align}
    &\bm{n\,.\,J} = \ell_c\big[\beta\,\phi - \alpha\,\phi_c\big], &\bm{x}\in\partial V, \label{eq:BC_flux_binding_unbinding}\\
    &\partial_t \phi_c = \beta\, \phi - \alpha\, \phi_c, &\bm{x}\in\partial V,  \label{eq:BC_phic}\\
    &\boldsymbol{n\,.\, \nabla}\phi=0,&\bm{x}\in\partial V,\label{eq:BC_wetting}
\end{align}
where $\phi_c$ is the volume fraction of surface-bound \textit{oskar} in a cortical layer of thickness $\ell_c$. $\beta$ is the uniform binding rate due to myosin-V activity, and $\alpha$ is the position-dependent unbinding rate due to kinesin-1 activity. Eq.~\ref{eq:BC_flux_binding_unbinding} balances the flux of \textit{oskar} at the cortex with the net exchange due to cortical binding and unbinding. According to Eqs.~\ref{eq:BC_flux_binding_unbinding} and \ref{eq:BC_phic}, the total amount of \textit{oskar} is conserved in the system:
\begin{align}
    \dfrac{\mathrm{d} }{\mathrm{d}t} \Big[ \ell_c \!\int_{\partial V} \!\!\! \phi_c(t)\, \mathrm{d}A+\int_{V} \!\! \phi(x,t)\, \mathrm{d}V\Big]=0.
\end{align}
 The unbinding rate varies along the cortex following the Patronin density distribution that reflects microtubule density gradients:
\begin{align}
    &\alpha(\bm{x}) = \alpha_{_\mathrm{SF}}\, I_{\mathrm{p}}(\bm{x}), &\bm{x}\in\partial V, 
\end{align}
where $I_{\mathrm{p}}$ is the normalized density distribution of Patronin, given in Fig.~3 of the main text. Eq.~\ref{eq:BC_phic} tracks the net accumulation of cortex-bound \textit{oskar} while Eq.~\ref{eq:BC_wetting} imposes a neutral wetting condition, ensuring no additional affinity between the \textit{oskar}-rich phase and the cortex. Together, Eqs.~\ref{eq:BC_flux_binding_unbinding} and \ref{eq:BC_wetting} are equivalent to: 
\begin{align}
&\boldsymbol{n\,.\,} \boldsymbol\nabla (\Delta \phi) = \dfrac{\ell_c}{\gamma}\Big[ \dfrac{\beta\, \phi -\alpha\, \phi_c}{\phi(1-\phi)} \Big], &&\bm{x}\in\partial V.\label{eq:BC_flux_gradlap_phi}
\end{align}
Nondimensionalizing the governing equations using $R$ as the length scale and $t_d=R^2/D$ as the time scale yields, in addition to $\chi$ and $\lambda$, the dimensionless groups
\begin{align}
    &A=\alpha_{_\mathrm{SF}}\, t_d, &&B=\beta\, t_d, &&\epsilon=\dfrac{\ell_c}{R}, &&\Gamma=\dfrac{\gamma\, t_d}{R^4}.
\end{align}
The nondimensional governing equations are:
\begin{align}
&\partial_t \phi =\psi + \Delta g(\phi) -\Gamma\,\bnabla \bcdot \big[ \phi(1-\phi) \, \bnabla \psi \big], &\bm{x}\in V,  \label{eq:transport_cf_nondim_imp}\\
&\psi=\Delta \phi,\\
&\boldsymbol{n\,.\,} \boldsymbol\nabla \psi = \dfrac{\epsilon}{\Gamma}\Big[\dfrac{B\, \phi-A\, \phi_c }{\phi(1-\phi)} \Big], &\bm{x}\in \partial V, \label{eq:BC_flux_gradlap_phi_nondim_imp}\\
 &\partial_t \phi_c = B\, \phi - A\, \phi_c.  &\bm{x}\in \partial V, \label{eq:BC_phic_nondim_imp}\\
 &\boldsymbol{n\,.\,\nabla}\phi=0, &\bm{x}\in \partial V. \label{eq:BC_wetting_nondim_imp}
\end{align}
For the analysis of photoconversion experiments in Sec.~\ref{sec:SI:photoconversion}, we take the limit $\chi, \Gamma\rightarrow0$ and $\lambda\rightarrow1$, in which Eqs.~\ref{eq:transport_cf_nondim_imp}--\ref{eq:BC_wetting_nondim_imp} reduce to simple diffusion in the bulk with the continuity of flux at the boundary: 
\begin{align}
&\partial_t \phi =\Delta \phi, &\bm{x}\in V,  \label{eq:transport_cf_nondim_limit}\\
&\boldsymbol{n\,.\,} \boldsymbol\nabla \phi = \epsilon\, \big[A\, \phi_c-B\, \phi\big], &\bm{x}\in \partial V, \label{eq:BC_flux_gradlap_phi_nondim_limit}\\
 &\partial_t \phi_c = B\, \phi - A\, \phi_c.  &\bm{x}\in \partial V. \label{eq:BC_phic_nondim_limit}
\end{align}

We solve Eqs.~\ref{eq:transport_cf_nondim_imp}--\ref{eq:BC_wetting_nondim_imp} numerically using the \texttt{Dedalus} spectral PDE solver in a spherical geometry of radius $R$~\cite{Burns_Dedalus_A_flexible_2020}. The linear terms are treated implicitly, while the nonlinear terms are treated explicitly. The initial condition is a uniform bulk volume fraction $\phi(\boldsymbol{x},0) = \phi_0$ with zero cortex-bound volume fraction $\phi_c(0) = 0$. The total amount of \textit{oskar} is thus set by the initial volume fraction $\phi_0$, a key parameter controlling the phase-separation behavior. The ratio of unbinding to binding rates, or the exchange equilibrium constant, $r_{\mathrm{ex}}=A/B = \alpha_{_\mathrm{SF}}/\beta$, determines the ratio of bulk to cortex-bound volume fractions at equilibrium, whereas the individual values of $A$ and $B$ affect only the time to reach equilibrium. The cortical layer thickness is measured to be $\ell_c \approx 10~\mathrm{\mu m}$ at the posterior pole, which yields $\epsilon \approx 0.25$ (with $R_{\mathrm{oocyte}} = R + \ell_c \approx 50~\mathrm{\mu m}$ at stage 9). The parameter $\Gamma$ controls the stability of the system against phase separation (increasing $\Gamma$ opposes phase separation) and sets the length scale of gradients in $\phi$. We set $\Gamma = B = 1$ and vary $A$ and $\phi_0$ to construct the phase diagram in Fig.~4 of the main text.

\vspace{1cm}

\section{Photoconversion of Staufen}\label{sec:SI:photoconversion}

In photoconversion experiments, an area of approximately $550~\mathrm{\mu m^2}$
is exposed to ultraviolet light. Staufen-MoxMaple3 switches irreversibly from
green to red fluorescence upon exposure, allowing a distinct population of
Staufen to be labeled and tracked simultaneously over time across three
distinct cortical regions: the anterior, mid-lateral, and posterior cortex. Red
fluorescence intensity is averaged over the box. We observe that variations in
fluorescence intensity are predominantly along the direction normal to the
cortex. Inspired by this observation, we use the model described in
Sec.~\ref{sec:SI:fitting_of_thermodynamic_model} in one dimension. The time
span of photoconversion is $10$~min, which is much shorter than that of
oogenesis from stage 7 to 9 ($\sim 12$~hours), where \textit{oskar} RNPs
aggregate and phase separate in the bulk. We therefore neglect the interactions
among \textit{oskar} RNPs, and between \textit{oskar} RNPs and the solvent, that
drive demixing in the bulk ($\chi = 0$). In the absence of strong
inhomogeneities in the bulk, we also neglect the corresponding gradient free
energy ($\kappa = 0$). The model then reduces to a diffusion equation on the
half-line, subject to cortical binding and unbinding at the boundary:
\begin{align}
    &\partial_t \phi = D\,\partial_x^2 \phi, &x \ge 0,\\
    &D\,\partial_x \phi = \ell_c\bigl[\beta\,\phi(0,t) - \alpha\,\phi_c(t)\bigr], &x = 0,
        \label{eq:phi_BC_flux_1D}\\
    &\partial_t \phi_c = \beta\,\phi(0,t) - \alpha\,\phi_c(t), &x = 0,
        \label{eq:phic_ode_1D}
\end{align}
where $\phi$ is the volume fraction of \textit{oskar} in the bulk and $\phi_c$
is the volume fraction of cortically bound \textit{oskar} within a layer of
thickness $\ell_c$. The constants $\beta$ and $\alpha$ are the binding and
unbinding rates (units of $\mathrm{s^{-1}}$). The initial conditions are
$\phi(x,0) = \phi_0\,[H(x+\ell_b) - H(x-\ell_b)]$ and $\phi_c(0)$, where $H(x)$ is the
Heaviside step function; on the half-line the initial bulk profile is a box of height
$\phi_0$ on $0 \le x \le \ell_b$. The exposed area $550~\mathrm{\mu m^2}$
corresponds to a square of side $\ell = 23~\mathrm{\mu m}$, with $\ell = \ell_b + \ell_c$.

We introduce the Laplace transform in time,
$\hat\phi(x,s) = \int_0^\infty e^{-st}\,\phi(x,t)\,dt$, and define
$q = \sqrt{s\,\ell_b^2/D}$. The diffusion equation becomes an ordinary
differential equation in $x$ in Laplace space with the initial condition as a source term, and
the cortical equation transforms algebraically:
\begin{align}
    &D\,\hat\phi_{xx} - s\,\hat\phi = -\phi(x,0),\\
    &\hat\phi_c(s) = \frac{\beta\,\hat\phi(0,s) + \phi_c(0)}{s + \alpha}.\label{eq:phic_solution_laplace}
\end{align}
The bulk volume fraction is a piecewise function,
\begin{equation} \label{eq:phi_solution_laplace}
\hat\phi(x,s) =
\begin{cases}
\dfrac{\phi_0}{s} + C_1\,e^{qx/\ell_b} + C_2\,e^{-qx/\ell_b}, & 0 \le x/\ell_b \le 1,\\[2mm]
C_3\,e^{-qx/\ell_b}, & x/\ell_b \ge 1,
\end{cases}
\end{equation}
Substituting \eqref{eq:phic_solution_laplace} and \eqref{eq:phi_solution_laplace} into the boundary condition
\eqref{eq:phi_BC_flux_1D}, and enforcing the continuity of $\hat\phi$ and
$\partial_x \hat\phi$ at $x = \ell_b$, we determine the coefficients:
\begin{align}
    &C_1 = -\frac{\phi_0}{2s}\,e^{-q},
    &C_2 = \hat\phi(0,s) - \frac{\phi_0}{s} + \frac{\phi_0}{2s}\,e^{-q},
    &&C_3 = \hat\phi(0,s) + \frac{\phi_0}{s}\bigl(\cosh q - 1\bigr),
\end{align}
where the boundary value is
\begin{equation} \label{eq:phi0_boundary_value}
\hat\phi(0,s) =
\frac{\phi_0\,\sqrt{D/s}\,\bigl(1 - e^{-q}\bigr)
      + \ell_c\,\dfrac{\alpha\,\phi_c(0)}{s + \alpha}}
     {\sqrt{Ds} + \ell_c\,\dfrac{\beta s}{s + \alpha}}. 
\end{equation}
Setting $\beta = 0$ and $\phi_c(0) = 0$ reduces the boundary value to
$\hat\phi(0,s) = \phi_0\,(1 - e^{-q})/s$, which inverts to
$\phi(0,t) = \phi_0\,\operatorname{erf}\!\bigl(\ell_b/(2\sqrt{Dt})\bigr)$, the
boundary value under a no-flux boundary condition. In this limit, $C_1 = C_2$ and the solution reduces to the
symmetric spreading of the box-shaped initial condition.

\begin{figure*}[tbhp]%
\centering
\includegraphics[width=0.45\linewidth]{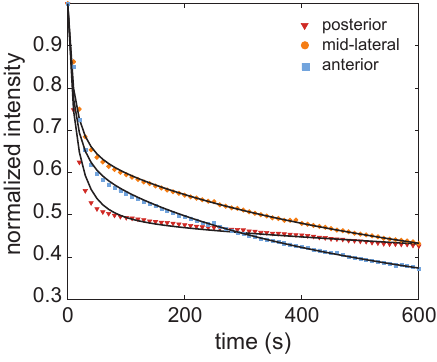}
\caption{Fluorescence intensity at three cortical positions in photoconversion experiments shown by markers ($n = 18$). Solid lines are the least-squares fits of the model with $(D/w^2,~\beta)=(0.0144,~0.0406)~(\mathrm{s^{-1}})$. The fitted unbinding rates are $(\alpha_p, \alpha_m, \alpha_a) = (0.00442, 0.0144, 0.0203)~(\mathrm{s^{-1}})$ and the bulk-thickness ratios are $((\ell_b/\ell)_p,~(\ell_b/\ell)_m,~(\ell_b/\ell)_a)=(0.847,~0.703,~0.793)$ at the posterior, mid-lateral, and anterior regions, respectively. }
\label{fig:photoconversion_fit_mean}
\end{figure*}

\begin{figure*}[tbhp]
\centering
\includegraphics[width=1.0\linewidth]{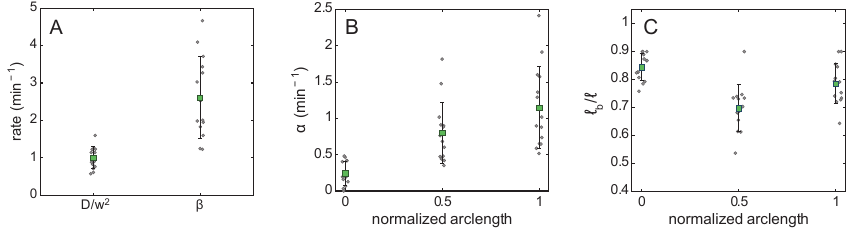}
\caption{Fit parameters of the model for photoconversion experiments ($n = 14$): (A) Diffusion rate $D/w^2$ and binding rate $\beta$ are constant across cortical positions, while (B) the unbinding rate $\alpha$ and (C) the bulk thickness ratio $\ell_b/\ell$ are position-dependent.}
\label{fig:photoconversion_fit_params}
\end{figure*}

We determine the box-averaged volume fraction 
of \textit{oskar} in a window of size $\ell$ over time. We compare its normalized value with the normalized fluorescent intensity obtained in photoconversion experiments: 
\begin{align}
    &\bar\phi(t) = \frac{1}{\ell}\big[\,\ell_c\,\phi_c(t)
        + \int_0^{\ell_b} \phi(x,t)\,\mathrm{d}x \big],
        \label{eq:phibar_def}\\
    &\bar\phi(0) = \frac{1}{\ell}\bigl[\,\ell_c\,\phi_c(0) + \phi_0\,\ell_b\,\bigr],\\
    &I_{\mathrm{obs}}(t) = \frac{\bar\phi(t)}{\bar\phi(0)}.
\end{align}
The bulk contribution to \eqref{eq:phibar_def} is
\begin{equation}
\int_0^{\ell_b} \hat\phi(x,s)\,\mathrm{d}x
   = \frac{\phi_0\,\ell_b}{s}
   + \frac{\ell_b}{q}\Bigl[\,C_1\bigl(e^{q} - 1\bigr)
                         + C_2\bigl(1 - e^{-q}\bigr)\,\Bigr],
\end{equation}
in Laplace space. The boundary value $\hat\phi(0,s)$ in
\eqref{eq:phi0_boundary_value} does not have an elementary inverse transform,
due to the $\sqrt{s}$ term that appears with the rational reaction factor
$\beta s/(s+\alpha)$ in the denominator. Therefore, $\phi(x,t)$, $\bar\phi(t)$,
and $I_{\mathrm{obs}}(t)$ are recovered by numerical Laplace inversion \cite{Abate_Valko_LaplaceInv_2004}. The rates $\alpha$,
$\beta$, $D/\ell^2$ and the thickness ratio $\ell_b/\ell$ are inferred by fitting
$I_{\mathrm{obs}}(t)$ to the measured intensity decay. Fig.~\ref{fig:photoconversion_fit_mean} shows the photoconversion intensity decay averaged over 18 oocytes, with the least-squares fit of the model overlaid as solid lines at each cortical position. Fig.~\ref{fig:photoconversion_fit_params} shows the distributions of fitted parameters across the same 18 oocytes, and Table~\ref{tab:photoconv_fit} lists the mean, standard deviation, and standard error of the mean for each parameter. Based on experimental measurements at $t = 0$ s, the cortical volume fraction exceeds the bulk volume fraction at the posterior pole. We therefore set $\phi_c(0)/\phi_0 = 2$ at the posterior pole, and $\phi_c(0)/\phi_0 = 1$ at the anterior pole and mid-lateral point.

\begin{table}[tbhp]
\centering
\caption{Fitted parameters from photoconversion experiments (mean $\pm$ SD across $n = 14$ experiments). Parameters $D/w^2$ and $\beta$ are constant across cortical positions, while $\alpha$ and $\ell_b/\ell$ are position-dependent.}
\label{tab:photoconv_fit}
\begin{tabular}{@{}llccc@{}}
\toprule
Parameter & Description & Mean & SD \\
\midrule
$D/w^2~(\mathrm{min}^{-1})$ & Diffusion rate & $1.01$ & $0.30$ \\
$\beta~(\mathrm{min}^{-1})$ & Binding rate & $2.61$ & $1.09$ \\
\addlinespace[5pt]
$\alpha_a~(\mathrm{min}^{-1})$ & Unbinding rate, anterior & $1.15$ & $0.57$ \\
$\alpha_m~(\mathrm{min}^{-1})$ & Unbinding rate, mid-lateral & $0.80$ & $0.42$ \\
$\alpha_p~(\mathrm{min}^{-1})$ & Unbinding rate, posterior & $0.24$ & $0.17$ \\
\addlinespace[5pt]
$(\ell_b/\ell)_a$ & Bulk fraction, anterior & $0.79$ & $0.073$ \\
$(\ell_b/\ell)_m$ & Bulk fraction, mid-lateral & $0.70$ & $0.084$ \\
$(\ell_b/\ell)_p$ & Bulk fraction, posterior & $0.84$ & $0.048$ \\
\bottomrule
\end{tabular}
\end{table}

The optical section thickness in spinning-disk confocal microscopy is much smaller than the lateral field of view, $w \approx 1~\mathrm{\mu m} \ll \ell \approx 23~\mathrm{\mu m}$. Consequently, \textit{oskar} RNPs diffuse out of the imaging plane faster than they traverse the field of view laterally. Therefore,  the observed fluorescence decay is dominated by out-of-plane diffusion. Taking the cytoplasm to be $\sim 1000$ times more viscous than water, we estimate the effective particle radius from the Stokes-Einstein relation together with the fitted diffusion rate in Table~\ref{tab:photoconv_fit}:
\begin{gather}
\frac{k_B T}{3\pi\,\mu\, a_{\mathrm{eff}}\, (w/2)^2} = 1~\mathrm{min}^{-1}, \\
a_{\mathrm{eff}} \approx 100~\mathrm{nm},
\end{gather}
which is comparable to the average \textit{oskar} RNP radius of $200~\mathrm{nm}$ reported in~\cite{bose_Cell_2022_liquid-to-solid_phase_transition}.

\FloatBarrier
\pagebreak
\section{Measuring the total \textit{oskar} mRNA in the oocyte} \label{sec:SI:total_osk}
\begin{figure*}[tbhp]
\centering
\includegraphics[width=.8\linewidth]{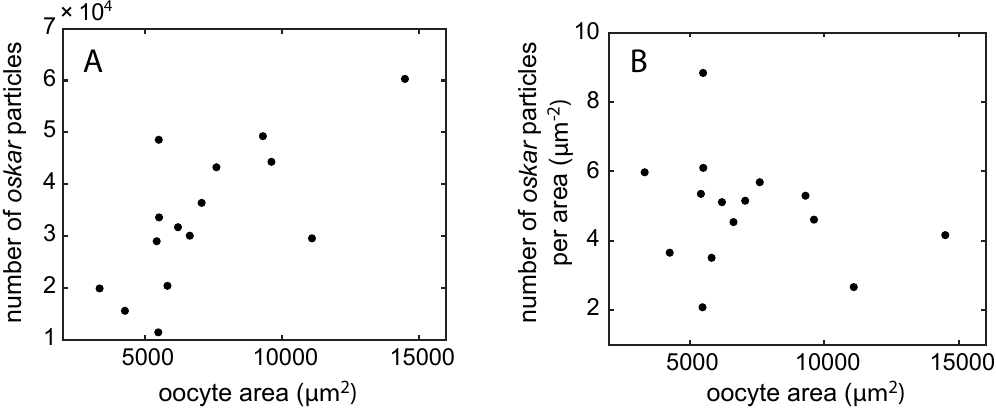}
\caption{(A) Estimated total number of \textit{oskar} mRNA particles plotted against oocyte area. Particle number was calculated by dividing the background-corrected total fluorescence intensity of each oocyte by the mean fluorescence intensity of a single cytoplasmic \textit{oskar} particle. (B) Estimated \textit{oskar} particle density, calculated as particle number per unit oocyte area, plotted against oocyte area. Each dot represents one oocyte.
}
\label{fig:osk_amount}
\end{figure*}

To estimate the number of \textit{oskar} mRNA granules at the posterior cap, \textit{oskar} was visualized by single-molecule inexpensive fluorescence in situ hybridization (smiFISH), and images were acquired as non-saturating $z$-stacks. For each oocyte, a single optical section containing clearly resolved cytoplasmic \textit{oskar} granules was selected from the $z$-stack to estimate the fluorescence intensity of individual particles. Single \textit{oskar} granules were detected using the Find Maxima function in Fiji/ImageJ, with the noise threshold adjusted for each image to match visually identifiable single granules. Because the $xy$ pixel size was approximately 200 nm, which is close to the expected size of \textit{oskar} granules, each local maximum detected by the Find Maxima function was treated as one \textit{oskar} granule. The fluorescence intensity of the pixels detected by the Find Maxima function was measured, and the average intensity of dim single particles was estimated by excluding the five lowest values and averaging the next 20 lowest mean-intensity values.
To quantify posterior-cap \textit{oskar}, the full \textit{oskar} $z$-stack was sum-projected. Background fluorescence was measured from three regions outside the egg chamber and averaged. The background-corrected total fluorescence intensity of the entire oocyte was calculated by subtracting the background signal from the oocyte signal. The estimated number of \textit{oskar} granules in the entire oocyte was then calculated by dividing this background-corrected oocyte total intensity by the average intensity of dim single cytoplasmic \textit{oskar} granules (Fig.\ref{fig:osk_amount}). The estimated \textit{oskar} mRNA granule abundance scales with oocyte size.

\FloatBarrier

\pagebreak

\section{Fitting of thermodynamic model to the experiments}\label{sec:SI:fitting_of_thermodynamic_model}

\begin{gather}
    f(\phi, T) = k_B\,T \bigg[ \frac{\phi}{\nu} \ln \phi  + \frac{(1-\phi)}{\nu_0} \ln (1-\phi) +\frac{\chi}{\nu_0} \, \phi(1-\phi)\bigg].\\
    \mu_{\mathrm{in}}=\mu_{\mathrm{out}}, \label{eq:equilibrium_mu}\\
    f_{\mathrm{in}}-\mu_{\mathrm{in}}\, \dfrac{\phi_{\mathrm{in}}}{\nu}=f_{\mathrm{out}}-\mu_{\mathrm{out}}\, \dfrac{\phi_{\mathrm{out}}}{\nu} + \dfrac{2\gamma}{R},\label{eq:equilibrium_Posm}
\end{gather}

Table~\ref{tab:parameters} lists the parameters used to determine the thermodynamic properties of \textit{oskar}--Staufen particles. The mean number of particles $N$ and mean oocyte cross-sectional area $A$ at stage 9 are measured in Fig.~\ref{fig:osk_amount}, and the mean particle radius $a$ is taken from \cite{bose_Cell_2022_liquid-to-solid_phase_transition}. Together with the sample thickness $h$, these yield estimates of the total particle concentration $c_0$, total volume fraction $\phi_0$, and the ratio $\nu/\nu_0$, where $\nu_0 = 18.015 \times 10^{-3}~\mathrm{M^{-1}}$ is the molar volume of water. 

\begin{figure*}[h]%
\centering
\includegraphics[width=1\linewidth]{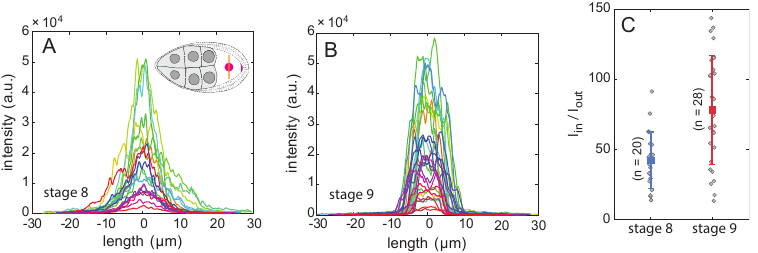}
\caption{Intensity profile along a line (thickness $7$--$10$ pixels, with $5.2~\mathrm{pixels} = 1~\mathrm{\mu m}$) perpendicular to the anterior--posterior axis, passing through the phase-separated droplet as shown in the inset,  under Patronin overexpression at (A) stage 8 and (B) stage 9. (C) Intensity ratio $I_{\mathrm{in}}/I_{\mathrm{out}}$ at each stage, with mean values and standard deviations $42\pm 20$ at stage 8 ($n=20$), and $78\pm39$ at stage 9 ($n=28$).
Error bars represent one standard deviation, and $n$ denotes the number
of oocytes analyzed. The $p$ value between the two groups is $p=0.0005$.}%
\label{fig:intensity_in_out_mutant}
\end{figure*}
\FloatBarrier

\

In Fig.~\ref{fig:intensity_in_out_mutant}A,B we measured the fluorescence intensity profile along a line through a phase-separated droplet at stages 8 and 9. For each oocyte, we first identified the $z$-slice of maximum \textit{oskar} intensity and drew a line of $10$ px width through the droplet, perpendicular to the anterior--posterior axis. We extracted the intensity profile along this line at the three adjacent $z$-slices. The intensity inside the droplet was estimated by averaging over pixels above $80\%$ of the maximum. Taking the center of this high-intensity region as a reference, the outside intensity was estimated by averaging over positions more than $20~\mathrm{\mu m}$ away, well beyond the droplet radius of $\lesssim 10~\mathrm{\mu m}$ at stage 9 (Fig.~\ref{fig:intensity_in_out_mutant}B). After subtracting the background intensity measured outside the oocyte, we computed the mean inside-to-outside intensity ratio $I_{\mathrm{in}}/I_{\mathrm{out}}=78.15$ at stage 9 (Fig.~\ref{fig:intensity_in_out_mutant}C).

Using this ratio, we determine the interaction parameter $\chi$ and the \textit{oskar} volume fractions inside and outside the droplet, $\phi_{\mathrm{in}}$ and $\phi_{\mathrm{out}}$, by solving \eqref{eq:equilibrium_mu} and~\eqref{eq:equilibrium_Posm} together with $I_{\mathrm{in}}/I_{\mathrm{out}} = \phi_{\mathrm{in}}/\phi_{\mathrm{out}}$, which assumes that intensity is proportional to volume fraction. A similar approach was used in~\cite{fritschLocalThermodynamicsGovern2021} to determine the thermodynamic parameters of P-granule condensation in \textit{Caenorhabditis elegans}.

\begin{table}[h]
\centering
\caption{List of parameters for the \textit{oskar}--Staufen particles.}
\label{tab:parameters}
\footnotesize
\setlength{\tabcolsep}{3pt}
\begin{tabular*}{\linewidth}{@{\extracolsep{\fill}} l cccccccc|ccc @{}}
\toprule
Parameter 
  & $N$ 
  & $a$ (nm) 
  & $A$ ($\mu$m$^2$) 
  & $h$ ($\mu$m)
  & $c_0$ ($\mu$M)  
  & $\phi_0$ 
  & $I_{\mathrm{in}}/I_{\mathrm{out}}$ 
  & $\nu/\nu_0$
  & $\chi$ 
  & $\phi_{\mathrm{in}}$ 
  & $\phi_{\mathrm{out}}$ \\
\midrule
Value 
  & $33544$ & $200$ & $7333$ & $30$ & $760$ & $0.0163\pm0.0054$ & $78.15$ & $1.12\times10^3$ & $0.55$ & $0.11$ & $0.0014$ \\
\bottomrule
\end{tabular*}
\end{table}

\pagebreak

\section{Asymmetric growth of the oocyte} \label{sec:SI:asymmetric_growth}
During stages 7 to 9, the oocyte grows up to fivefold in length (pole-to-pole distance from $\sim 20~\mathrm{\mu m}$ to $\sim 100~\mathrm{\mu m}$ over $\sim 12$ hours), predominantly from the anterior side. This corresponds to a characteristic growth speed $u_g \approx 1.9~\mathrm{nm/s}$. Using the effective particle radius $a_{\mathrm{eff}} \approx 100~\mathrm{nm}$ determined from the photoconversion experiments in Sec.~\ref{sec:SI:photoconversion} and an effective cytoplasmic viscosity $1000$ times that of water, we estimate the diffusive transport speed of \textit{oskar} RNPs across the oocyte to be $u_d = k_B T/(2\pi \mu \,a_{\mathrm{eff}}\, R_{\mathrm{oocyte}}) \approx 0.13~\mathrm{nm/s}$. Since diffusion is much slower than growth ($u_g/u_d>10$), we incorporate the effect of asymmetric growth in two discrete steps. We first solve the problem in a smaller geometry representing the stage-7 oocyte, and then use the resulting steady state as the initial condition for a separate simulation in a geometry five times larger. The details are described below.

The simulations are performed in an axisymmetric spherical domain $V_f$ of radius $R_f = 5$ about the $z$-axis. The initial condition is constructed from the steady state of a separate simulation on a smaller spherical domain $V_i$ of radius $R_i = 1$ with uniform binding and unbinding. The main challenge is that the numerical grids of the two domains do not coincide: they have different spatial resolutions and their centers are offset. Here we describe how the solution on $V_i$ is mapped onto the $V_f$ grid.

The domain $V_i$ is placed at the posterior side of $V_f$, its center displaced along the anterior--posterior ($z$) axis by $z_{\mathrm{shift}} = R_f - R_i = 4$, so that the two share a common pole (Fig.~5 of the main text). We define radial and polar coordinates measured from the center of $V_i$,
\begin{align}
    r(\bm{x}) &= \sqrt{x^2 + y^2 + (z - z_{\mathrm{shift}})^2}, \qquad \bm{x} \in V_f,\\
    \theta(\bm{x}) &= \arccos\!\left(\frac{z - z_{\mathrm{shift}}}{r}\right).
\end{align}
For every $\bm{x} \in V_f$, we evaluate the initial volume fraction from the steady-state $V_i$ solution at $(r, \theta)$ directly from its spectral representation,
\begin{equation}
  \phi_{f0}(r,\theta) =
  \begin{cases}
    \phi_i(r, \theta, t \to \infty), & r(\bm{x}) \le R_i,\\[2pt]
    \phi_i(R_i, \theta, t \to \infty), & r(\bm{x}) > R_i,
  \end{cases}
\end{equation}
where the azimuthal angle is immaterial in the axisymmetric geometry. This mapping is exact up to spectral truncation error, even though the two simulations use non-overlapping grids with different resolutions. Outside $V_i$, $\phi$ is set to a constant value $\phi_\infty$. To connect the interior and exterior values smoothly, we use a hyperbolic-tangent weight of half-width $\delta = 0.25$,
\begin{equation}
    B(r) = \tfrac{1}{2}\big[1 - \tanh\!\big((r - R_i)/\delta\big)\big],
\end{equation}
so that the initial condition is
\begin{equation}
    \phi(\bm{x}, 0) = B(r)\,\phi_{f0}(r, \theta) + \big[1 - B(r)\big]\,\phi_\infty, \qquad \bm{x} \in V_f.
\end{equation}
The weight function departs from $0$ and $1$ only within a thin shell around $r = R_i$, so that both the interior solution and the exterior value $\phi_\infty$ are recovered away from this boundary. In the simulation of Fig.~5, we set $\phi_\infty = 0.0107$, giving an average volume fraction $\phi_0 = |V_f|^{-1} \int_{V_f} \phi(\bm{x}, 0) \, \mathrm{d}v = 0.011$. Finally, we initialize the cortical field in local equilibrium with $\phi$ at the boundary ($\partial_t \phi_c = 0$),
\begin{equation}
    \phi_c(\bm{x}, 0) = \frac{\beta}{\alpha(\bm{x})}\, \phi(\bm{x}, 0), \qquad \bm{x} \in \partial V_f.
\end{equation}

\bibliographystyle{unsrt-keepcase}
\bibliography{Refs}